\documentclass[hyper]{JHEP3}
\usepackage{amsmath,amsfonts}
\usepackage{epsfig}
\usepackage{cite}

\title{Affine $SU(N)$ algebra from wall-crossings}
\author{Takahiro Nishinaka$^a$\footnote{nishinak [at] post.kek.jp}\; and\; Satoshi Yamaguchi$^b$\footnote{yamaguch [at] het.phys.sci.osaka-u.ac.jp}\\
$^a$ High Energy Accelerator Research Organization (KEK),\\
 ~ Tsukuba, Ibaraki 305-0801, Japan\\
$^b$ Department of Physics, Graduate School of Science, Osaka University,\\
 ~ Toyonaka, Osaka 560-0043, Japan\\
}
\abstract{We study the relation between the instanton counting on ALE spaces and the BPS state counting on a toric Calabi-Yau three-fold. We put a single D4-brane on a divisor isomorphic to $A_{N-1}$-ALE space in the Calabi-Yau three-fold, and evaluate the discrete changes of BPS partition function of D4-D2-D0 states in the wall-crossing phenomena. In particular, we find that the character of affine $SU(N)$ algebra naturally arises in wall-crossings of D4-D2-D0 states. Our analysis is completely based on the wall-crossing formula for the $d=4, \mathcal{N}=2$ supersymmetric theory obtained by dimensionally reducing the Calabi-Yau three-fold. 
}
\preprint{KEK-TH 1476
\\
OU-HET 717}

\begin{document}

\bibliographystyle{JHEP}   



\section{Introduction}

The cohomology of the moduli space of instantons on $A_{N-1}$ ALE space is known to be acted on by affine $SU(N)$ algebra \cite{Nakajima}. As a result, the instanton partition function on $A_{N-1}$-ALE space is given by the character of a representation of the algebra. In particular, for $U(1)$ instantons, it is given by the level one character of affine $SU(N)$ algebra:
\begin{eqnarray}
 \chi_r^{\widehat{su}(N)_1}(q,Q) = \frac{1}{\eta(q)^{N-1}}\;\Theta_{r}^{{\widehat{su}(N)_1}}(q,Q),\label{eq:instanton-A_N}
\end{eqnarray}
where $q$ and $Q$ denote the chemical potentials for instantons and magnetic fluxes on $N-1$ blowup two-cycles, respectively.\footnote{For the precise definition of the theta function $\Theta_{r}^{\widehat{su}(N)_1}(q,Q)$, see equation \eqref{eq:Theta_function}.} The subscript $r=0,1,\cdots,N-1$ labels the representation of the algebra. This fact was shown by careful treatments of the (generalized) ADHM constraints for instantons on ALE spaces \cite{Nakajima}.

As is well-known, the ADHM constraints for instantons are equivalent to the BPS condition of D4-D0 systems in the field theory limit \cite{Witten:1994tz, Witten:1995gx, Douglas:1995bn, Douglas:1996uz, Douglas:1996sw}. Here, the field theory limit means the limit where $\alpha'$ and $g_s$-corrections are negligible. In this limit, BPS D0-branes are regarded as instantons in the field theory on D4-branes, where the temporal direction is appropriately reduced by the compactification on a circle. If there are also D2-branes wrapped on some compact two-cycle embedded in the D4-brane,\footnote{Here, we assume the two-cycle is transverse to the temporal direction.} they can be seen as non-trivial magnetic fluxes. 
Thus, in the field theory limit, counting the number of instantons and magnetic fluxes on a D4-brane is equivalent to counting BPS D2-D0 states on the D4-brane.
 This implies that the above instanton partition function \eqref{eq:instanton-A_N} is equivalent to the BPS partition function of D4-D2-D0 bound states on $A_{N-1}$ ALE space, {\em if there are no $\alpha'$ and $g_s$-corrections.}\footnote{In this paper, BPS partition function is a generating function of the degeneracy (or index) of stable BPS states.}

In this paper, we embed this D4-D2-D0 system on $A_{N-1}$ ALE space into a toric Calabi-Yau three-fold, and study the relation between the instanton counting on the ALE space and the BPS state counting on the Calabi-Yau three-fold. Note that wrapped D-branes on a Calabi-Yau three-fold can be seen as BPS particles in $d=4, \mathcal{N}=2$ supersymmetric gauge theory obtained by dimensionally reducing the Calabi-Yau. Since the BPS index of $d=4,\mathcal{N}=2$ theory is independent of the hypermultiplet moduli, we find that {\em there is no $g_s$-correction at all to the BPS index of our D4-D2-D0 states.} On the other hand, there might be non-vanishing $\alpha'$-corrections which modify the relation between the BPS partition function of D-branes and the instanton partition function on the ALE space. In general, the $\alpha'$-corrections are controlled by dimensionless parameters
\begin{eqnarray}
\frac{\alpha'}{R_i^2},
\end{eqnarray}
where $R_i$ denotes the radius of $i$-th compact two-cycle in the Calabi-Yau three-fold. Hence, if we take the {\em large radii limit} of $R_i\to \infty$, all the $\alpha'$-corrections are completely suppressed. This implies that, in the large radii limit, the BPS partition function of our D4-D2-D0 states on the Calabi-Yau coincides with the instanton partition function on ALE space.

Now, let us consider what happens if the radius of some compact cycle becomes finite or small. The radii of compact cycles belong to K\"ahler moduli of the Calabi-Yau, and changing the K\"ahler moduli in general modifies the BPS conditions of the D4-D2-D0 bound states. This implies that some BPS bound states of D4-D2-D0 branes might be unstable or newly appear in the spectrum when we change the radii of the cycles, which gives rise to discontinuous changes in the BPS partition function. This is called the {\em wall-crossing phenomenon} of BPS states.\footnote{For recent progress in the study of wall-crossing phenomena of D4-D2-D0, see \cite{Diaconescu:2007bf, Jafferis:2007ti, Andriyash:2008it, Manschot:2009ia, Szabo:2009vw, Manschot:2010xp, Nishinaka:2010qk, Nishinaka:2010fh, Alim:2010cf, Nishinaka:2011sv, Nishinaka:2011pd}.} Since the instanton partition function has no moduli dependence, this implies that the BPS partition function of our D4-D2-D0 system ceases to be equivalent to the instanton partition function on $A_{N-1}$ ALE space, when the wall-crossings occur. In this paper, we evaluate such discontinuous changes in the BPS partition function, by using the wall-crossing formula recently proposed in \cite{Denef:2007vg, Kontsevich-Soibelman, Kontsevich:2009xt}.

The most interesting fact we reveal in this paper is that the affine $SU(N)$ character \eqref{eq:instanton-A_N} naturally arises in the wall-crossing phenomena of D4-D2-D0 bound states. More precisely, we show that the character of $\widehat{su}(N)$ can be obtained from the instanton partition function on $\mathbb{C}^2$ via the wall-crossings. For that, we ``add'' a dummy compact two-cycle to our toric Calabi-Yau three-fold that has a divisor isomorphic to $A_{N-1}$-ALE space. If we take a non-compact limit of the dummy cycle, we should recover the original field theory description of our D4-D2-D0 branes on $A_{N-1}$-ALE space. Instead, we can consider the flop transition with respect to the dummy cycle, which changes the topology of the divisor wrapped by the D4-brane. After the flop, the large radii limit no longer gives the affine $SU(N)$ character \eqref{eq:instanton-A_N} but rather leads to a different field theory description of our D4-D2-D0 states. This is because now the divisor wrapped by the D4-brane is topologically different from ALE space. By considering more flops, we can finally make our divisor isomorphic to $\mathbb{C}^2$. Then, the large radii limit now gives the $U(1)$ instanton partition function on $\mathbb{C}^2$. What is important here is that these different large radii limits should be connected by wall-crossings. Thus, we can expect that the wall-crossings involving the flop transitions {\em interpolate} the instanton partition functions on ALE spaces and $\mathbb{C}^2$.  In this paper, we explicitly verify this, that is, the affine $SU(N)$ character can be obtained from the instanton partition function on $\mathbb{C}^2$, by considering the wall-crossings of D4-D2-D0. Our analysis is completely based on the wall-crossing formula for $d=4,\mathcal{N}=2$ supersymmetric theories. So our result implies that the wall-crossing formula {\em knows} about the instantons on ALE spaces.

The rest of this paper is organized as follows. In section \ref{sec:wall-crossing}, we explain our configuration of D4-D2-D0 branes on a Calabi-Yau three-fold. We also discuss the general feature of the wall-crossings of our D-branes. The wall-crossing formula of $d=4, \mathcal{N}=2$ supersymmetric theories is briefly reviewed. In section \ref{sec:affine_SU(2)}, we explain how the wall-crossings of D4-D2-D0 states interpolate the instanton partition functions on $A_1$-ALE space and $\mathbb{C}^2$. We explicitly obtain the affine $SU(2)$ character from the instanton partition function on $\mathbb{C}^2$ via the wall-crossing phenomena. In section \ref{sec:affine_SU(N)}, we generalize the argument in the previous section to $A_{N-1}$-ALE spaces. The result implies that the wall-crossing formula knows about the instantons on $A_{N-1}$-ALE spaces. In section \ref{sec:discussions}, we conclude and give some future directions. We have several appendices for some basic facts about the BPS indices, flop transition, Gopakumar-Vafa invariants and the affine $SU(N)$ character. Our notation is also fixed in appendix \ref{app:notation}. 

\section{Wall-crossing of D4-D2-D0 and ALE instantons}
\label{sec:wall-crossing}

\subsection{Embedding in a toric Calabi-Yau}

Let us consider a toric Calabi-Yau three-fold $X$ whose webdiagram is depicted as in figure \ref{fig:toric-A_{N-1}}. 
\begin{figure}
\begin{center}
\includegraphics[width=6cm]{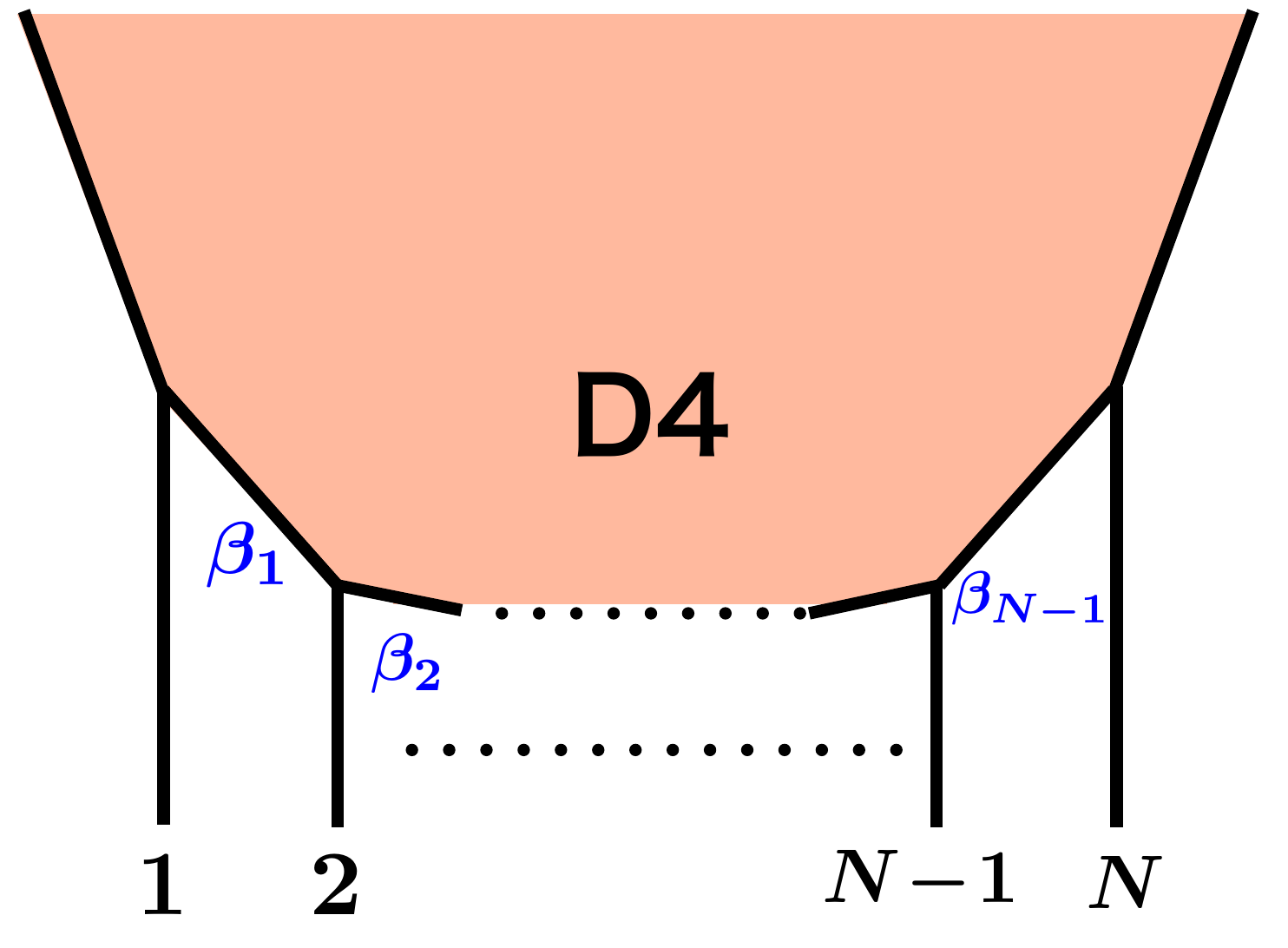}
\caption{The toric webdiagram of a Calabi-Yau three-fold that has $A_{N-1}$ ALE space as its divisor. The shaded region is a projection of the divisor. The diagram has $N+2$ external edges and $N-1$ internal lines, where the latter corresponds to $N-1$ blowup cycles in the resolved $A_{N-1}$ ALE space. $N$ numbered external edges are mutually parallel. When a single D4-brane is wrapped on the shaded divisor, the low-energy effective theory on the D4-brane is topologically twisted $\mathcal{N}=4\; U(1)$ gauge theory on $A_{N-1}$ ALE space.}\label{fig:toric-A_{N-1}}
\end{center}
\end{figure}
It has $N+2$ external edges and $N-1$ internal lines, where $N$ of the former are parallel to each other. This Calabi-Yau $X$ has a non-compact subspace (or divisor) which is isomorphic to $A_{N-1}$ ALE space, whose projection onto the toric base is the shaded region in figure \ref{fig:toric-A_{N-1}}. The $N-1$ internal lines in the diagram express $N-1$ compact two-cycle in $X$, which correspond in the divisor to $N-1$ blowup cycles of the resolved $A_{N-1}$ ALE space. We wrap a single D4-brane on the divisor, and consider D2-branes wrapped on compact two-cycles of $X$ as well as D0-branes localized in $X$.
As mentioned before, if we take the large radii limit then the BPS partition function of the D4-D2-D0 states becomes equivalent to the instanton partition function on ALE space. Note here that, since our D4-brane is now wrapped on a curved divisor, we need some topological twist to keep the supersymmetry. When the divisor is a holomorphic divisor of a Calabi-Yau three-fold, the low energy effective theory on the D4-brane is the Vafa-Witten theory \cite{Vafa:1994tf}. Therefore, the above instanton partition function on a D4-brane is that of the Vafa-Witten theory on ALE space.

When the K\"ahler moduli of the Calabi-Yau $X$ move away from the large radii limit, then the BPS partition function might be changed due to the wall-crossing phenomena.
Since the BPS indices are integer-valued, the BPS partition function could change only discontinuously. The subspaces in the moduli space where such discontinuous changes occur are called ``walls of marginal stability.'' As will be explained in the next subsection, these walls are generally real codimension one subspaces in the moduli space. So the moduli space is divided into chambers surrounded by the walls, and in each such chamber the BPS indices are exactly constant. At each wall of marginal stability, some BPS state could marginally decay into other BPS states, because otherwise the BPS indices cannot be changed.\footnote{For example, even if some BPS states pair up to form a non-BPS state, it gives no change in the BPS index. This is similar to the Witten index for supersymmetric vacua. The indices could be changed only when some short multiplet decays into other short multiplets or short multiplets form a new short multiplet.} Thus, we can identify the walls of marginal stability by considering all possible decay channels of the BPS states of interest. For our D4-D2-D0 states, it turns out in the next subsection that the only possible decay channels are separations of some D2-D0 fragments.

\subsection{Wall-crossing formula}
\label{subsec:wall-crossing_formula}

We here investigate possible decay channels for our D4-D2-D0 bound states in order to study the wall-crossing phenomena. We will also briefly review the relevant wall-crossing formula proposed in \cite{Denef:2007vg, Kontsevich-Soibelman}, and apply it to our D4-D2-D0 system.

Our notation is summarized in appendix \ref{app:notation}. We denote by $\mathcal{D}\in H^2(X)$ the charge of our non-compact D4-brane. There are $N-1$ independent unit D2-brane charges $\beta_i\in H^4(X)$ for $i=1,2,\cdots,N-1$. A unit D0-brane charge is denoted by $-dV\in H^6(X)$ which is a normalized volume form of the Calabi-Yau $X$ (up to sign), that is, $\int_X dV = 1$. With these definitions, the total charge $\Gamma$ of our D4-D2-D0 bound states can be written as
\begin{eqnarray}
\Gamma &=& \mathcal{D} + k^i\beta_i - ldV,\label{eq:charge}
\end{eqnarray}
where the integers $k^i$ and $l$ are D2 and D0-brane charges, respectively. 

Let $\mathcal{P}^i\in H^2(X)$ be the basis of compact harmonic two-forms on $X$, where $i$ runs over $i=1,2,\cdots,N-1$. We take these basis two-forms $\mathcal{P}^i$ so that
\begin{eqnarray}
\int_X \mathcal{P}^i\wedge \beta_j &=& \delta^i_j.
\end{eqnarray}
 The K\"ahler two-form $t$ of the Calabi-Yau $X$ can be written as
\begin{eqnarray}
 t &=& z_i\mathcal{P}^i + \Lambda e^{i\varphi}\mathcal{P}',
\end{eqnarray}
where $z_i$ parameterizes the complex $(N-1)$-dimensional K\"ahler moduli space. The additional parameter $\Lambda e^{i\varphi}$ denotes the K\"ahler parameter for non-compact cycles, and $\mathcal{P}'\in H^2(X)$ is harmonic two-form for the non-compact cycles. In the final result we should take the local limit $\Lambda\to\infty$, but even in the limit the result depends on $\varphi$. The phase $\varphi$ roughly means the ``ratio'' of the sizes and B-fields of the non-compact cycles. In the rest of this paper, we will keep the non-vanishing $\varphi$ fixed, and regard $z_i$ as moduli parameters. This extension of the moduli space was first given in \cite{Jafferis:2008uf}, and also used in \cite{Nishinaka:2010qk, Nishinaka:2010fh}. As in \cite{Nishinaka:2010qk, Nishinaka:2010fh}, we fix $\varphi$ so that $\pi/4 < \varphi < \pi/2$. Since $\Lambda e^{i\varphi}$ expresses the K\"ahler parameter for a non-compact cycle, $\mathcal{P}'$ should satisfy
\begin{eqnarray}
 \int_X \mathcal{P}'\wedge \beta_i &=& 0
\end{eqnarray}
for all $i = 1,2,\cdots,N-1$.

The central charge $Z(\Gamma)$ of a BPS state is a linear function of its charge $\Gamma$ \cite{Witten:1978mh}, which in the large radii limit can be written as (See appendix \ref{app:notation})
\begin{eqnarray}
Z(\Gamma) &=& -\int_X \Gamma\wedge e^{-t} \;\sim\; -\frac{c_4}{2}\Lambda^2e^{2i\varphi} + m^iz_i + n,\label{eq:central_charge}
\end{eqnarray}
up to a real positive prefactor.\footnote{In this paper, we use the expression of the central charges only to identify the locations of walls of marginal stability. As shown below, the locations of the walls depend only on the phases of the central charges and are independent of such a real positive prefactor. So we omit it. In literature, the expression \eqref{eq:central_charge} is sometimes called the holomorphic central charge.}
Here we defined a real constant $c_4 = \int_X \mathcal{D}\wedge \mathcal{P}'\wedge \mathcal{P}'$.
If the moduli move away from the large radii limit, then the geometric picture ceases to be valid due to the $\alpha'$-corrections. Thus the D2-brane and D0-brane contribution $m^iz_i+n$ may not be valid any more.  However, we can choose the normalization of the central charge and the coordinates of the moduli space so that the D2-brane and D0-brane contribution $m^iz_i+n$ is always valid as follows.  First, we choose the normalization of the central charge $Z(\Gamma)$ by the K\"ahler transformation in the moduli space, so that the central charge of a D0-brane is $1$. Next, we can redefine the moduli $z_i$ by a holomorphic coordinate transformation so that the central charge of the wrapped D2-branes is equal to $m^iz_i$.
Even after this redefinition, we can still identify ${\rm Im}\,z_i$ and ${\rm Re}\,z_i$ as the size and B-field of the $i$-th two-cycle in the limit of ${\rm Im}\,z_i\to \infty$. Hereafter, we use these redefined moduli parameters $z_i$.

One might think that the central charge of the D4-brane $-\frac{c_4}{2}\Lambda^2e^{2i\varphi}$ should also be modified when the moduli move away from the large radii limit. However, since our D4-brane is non-compact, the $\alpha'$-corrections are always small compared to the leading term $\int_X\mathcal{D}\wedge t\wedge t \sim -\frac{c_4}{2}\Lambda^2e^{2i\varphi}$. Thus, we can always use the expression \eqref{eq:central_charge} as long as we appropriately redefine $z_i$ as mentioned above.\footnote{If one considers a Calabi-Yau manifold with a compact 4-cycle, there is a $\alpha'$ corrections to the central charge of a D4-brane wrapping the 4-cycle. This correction cannot be absorbed by any redefinition.}

Now, we want to identify the walls of marginal stability for our BPS D4-D2-D0 states. To identify the walls, we have to know the possible decay channels of our BPS states with charge \eqref{eq:charge}. Suppose a BPS state with charge \eqref{eq:charge} decays into other two BPS states with charges $\Gamma_1$ and $\Gamma_2$, respectively. For generic values of the moduli, this BPS decay is forbidden by supersymmetry and conservation laws. From the charge conservation $\Gamma = \Gamma_1 + \Gamma_2$, it follows that $Z(\Gamma) = Z(\Gamma_1) + Z(\Gamma_2)$. Thus, we have the triangle inequality $\left|Z(\Gamma)\right|\leq \left|Z(\Gamma_1)\right| + \left|Z(\Gamma_2)\right|$. Since the mass of a BPS state is equal to the absolute value of its central charge, this implies that the BPS decay $\Gamma\to\Gamma_1 + \Gamma_2$ could occur only if $\arg{Z(\Gamma_1)} = \arg{Z(\Gamma_2)}$. Recall here the central charge $Z(\Gamma)$ implicitly depends on the moduli $z_i$. Thus, this condition can be solved for $z_i$, which gives us a real codimension one subspace in the moduli space. This subspace is called the wall of marginal stability, where the BPS indices for charge $\Gamma$ could be changed discontinuously.

For our case, we can generally consider the following charges of $\Gamma_1$ and $\Gamma_2$:
\begin{eqnarray}
\Gamma_1 &=& -a + (1-b)\mathcal{D} + (k^i- m^i)\beta_i - (l-n)dV,
\\
\Gamma_2 &=& a + b\mathcal{D} + m^i\beta_i - ndV,
\end{eqnarray}
where the integers $a,b,m^i$ and $n$ denote the D6, D4, D2 and D0-brane charges of $\Gamma_2$. The corresponding central charges are evaluated as
\begin{eqnarray}
 Z(\Gamma_1) &=& -\frac{ac_6}{6}\Lambda^3e^{3i\varphi} - \frac{(1-b)c_4}{2}\Lambda^2e^{2i\varphi} + (k^i-m^i)z_i + (l-n),
\\
 Z(\Gamma_2) &=& \frac{ac_6}{6}\Lambda^3e^{3i\varphi} - \frac{bc_4}{2}\Lambda^2e^{2i\varphi} + m^iz_i + n,
\end{eqnarray}
where $c_6 = \int_X\mathcal{P}'\wedge \mathcal{P}'\wedge \mathcal{P}'$.
We associate walls of marginal stability with these decay channels. They are defined as a subspace where $\arg[Z(\Gamma_1)] = \arg[Z(\Gamma_2)]$ is satisfied. It will turn out that there are only a single type of walls in the finite $z_i$ region. First, if we assume $a\neq 0$, then $Z(\Gamma_1)$ and $Z(\Gamma_2)$ are dominated by the D6 and $\overline{\rm D6}$-brane contributions $\pm \frac{ac_6}{6}\Lambda^3e^{3i\varphi}$ and never aligned. In fact they are always {\em anti}-aligned, and the corresponding BPS decay is forbidden in the local limit due to the energy conservation. So we can assume $a=0$. Next, let us consider the case of $b\neq 0,1$. In that case, $Z(\Gamma_1)$ and $Z(\Gamma_2)$ are dominated by the D4 (or $\overline{\rm D4}$) contribution and again anti-aligned. Therefore, we can concentrate on the case of $a =b=0$ without loss of generality. The possible decay channels of our BPS D4-D2-D0 states are then only those for
\begin{eqnarray}
 \Gamma_1 = \mathcal{D} + (k^i-m^i)\beta_i - (l-n)dV,\qquad \Gamma_2 = m^i\beta_i - ndV,\label{eq:possible_decays}
\end{eqnarray}
that is, the separations of D2-D0 fragments.

Now, we review the wall-crossing formula proposed in \cite{Denef:2007vg, Kontsevich-Soibelman}. The formula relevant for our analysis is the so-called ``semi-primitive formula''\cite{Denef:2007vg}. The semi-primitive formula is applicable if the wall-crossing is associated with a decay of $\Gamma \to \Gamma_1 + k\Gamma_2'$ for some positive integer $k\in\mathbb{N}$ and primitive charges $\Gamma_1,\,\Gamma_2'$. Here, we say a charge $\Gamma$ is ``primitive'' if $\Gamma$ has no positive integer $k$ that can divide out $\Gamma$. In our case, $k$ is the maximal common divisor of $m^i$ and $n$, and $\Gamma_2'=\Gamma_2/k$. Suppose the moduli cross a wall of marginal stability associated with a BPS decay channel $\Gamma\to\Gamma_1+k\Gamma_2'$ for some $k\in\mathbb{N}$. The central charges for $\Gamma_1$ and $\Gamma_2'$ are aligned at the wall: $\arg Z(\Gamma_1) = \arg Z(\Gamma_2')$. This implies that all the walls associated with $\Gamma \to \Gamma_1 + j\Gamma_2'$ for $j\in \mathbb{N}$ are, if they exist, also crossed simultaneously. It is for this reason that the semi-primitive wall-crossing formula is written in terms of the primitive $\Gamma_2'$ rather than $\Gamma_2 = k\Gamma_2'$.  Since the central charges $Z(\Gamma_1)$ and $Z(\Gamma_2')$ are aligned at the wall, the ordering of the central charge phases are generically reversed at the wall-crossing.\footnote{Otherwise, there is no change in the BPS indices.} We here assume $\arg Z(\Gamma_1)>\arg Z(\Gamma_2')$ before the wall-crossing and $\arg Z(\Gamma_1)<\arg Z(\Gamma_2')$ after the wall-crossing. Then, the wall-crossing formula says that through the wall-crossing the BPS partition function (defined in eq.~\eqref{def-partition-function1}) has a multiplication of the form
\begin{eqnarray}
\mathcal{Z}(q,Q) \to \mathcal{Z}(q,Q)\times\prod_{j=1}^\infty\left(1+ (-1)^{j\langle\Gamma_2',\Gamma\rangle}q^{jn'}\prod_{i=1}^{N-1}Q_i^{jm'^{i}}\right)^{j\langle \Gamma_2',\Gamma\rangle\Omega(j\Gamma_2')},\label{eq:wall-crossing_formula}
\end{eqnarray}
where $q$ and $Q_i$ denote the chemical potentials for D0 and D2-branes, and $\langle\Gamma,\Gamma'\rangle$ is the Dirac-Schwinger-Zwanziger intersection product (See appendix \ref{app:notation}). The integers $m'^i$ and $n'$ are the D2 and D0-brane charges of the primitive charge $\Gamma_2'$.\footnote{That is, $m'^i = m^i/k,\, n' = n/k$ where $k$ is the maximal common divisor of $m^i$ and $n$.}

We can easily apply this wall-crossing formula to our problem. Recall that the topology of our Calabi-Yau three-fold is described by the toric diagram depicted in figure \ref{fig:toric-A_{N-1}}. In particular, the intersection product between two-cycles and the divisor wrapped by the D4-brane can be read off as
\begin{eqnarray}
\left\langle \beta_i,\mathcal{D}\right\rangle &=& 0.
\end{eqnarray}
This implies that $\langle j\Gamma_2',\Gamma\rangle$ vanishes for all $\Gamma_2'$ associated with the decay channel \eqref{eq:possible_decays}. Therefore, {\em our D4-D2-D0 states have no wall-crossing with respect to the moduli for $N-1$ blowup cycles of the ALE space.} 
This is very simple but not interesting result. 

However, if we ``add'' an another compact two-cycle at an edge of the toric webdiagram, we can introduce non-vanishing effects of wall-crossings to our D4-D2-D0 system. In section \ref{sec:affine_SU(2)} and \ref{sec:affine_SU(N)}, we will find that such an ``adding'' gives us a rich class of interesting wall-crossing phenomena, and reveals the relation between instanton counting on the ALE spaces and BPS counting on a toric Calabi-Yau three-fold. In the next section, in order to make our idea clear, we concentrate on the simplest case of $N=2$, that is, the relation between instantons on $A_1$-ALE space and the D4-D2-D0 wall-crossings. In section \ref{sec:affine_SU(N)}, we will generalize it to the $A_{N-1}$-ALE case.

\section{Affine $SU(2)$ algebra from wall-crossings}
\label{sec:affine_SU(2)}

In this section, we examine the relation between the instanton counting on $A_1$-ALE space and the D4-D2-D0 wall-crossings. In particular, we will obtain the affine $SU(2)$ character from the instanton partition function on $\mathbb{C}^2$ by considering the wall-crossings of D4-D2-D0 states.

\begin{figure}
\begin{center}
\includegraphics[width=5.5cm]{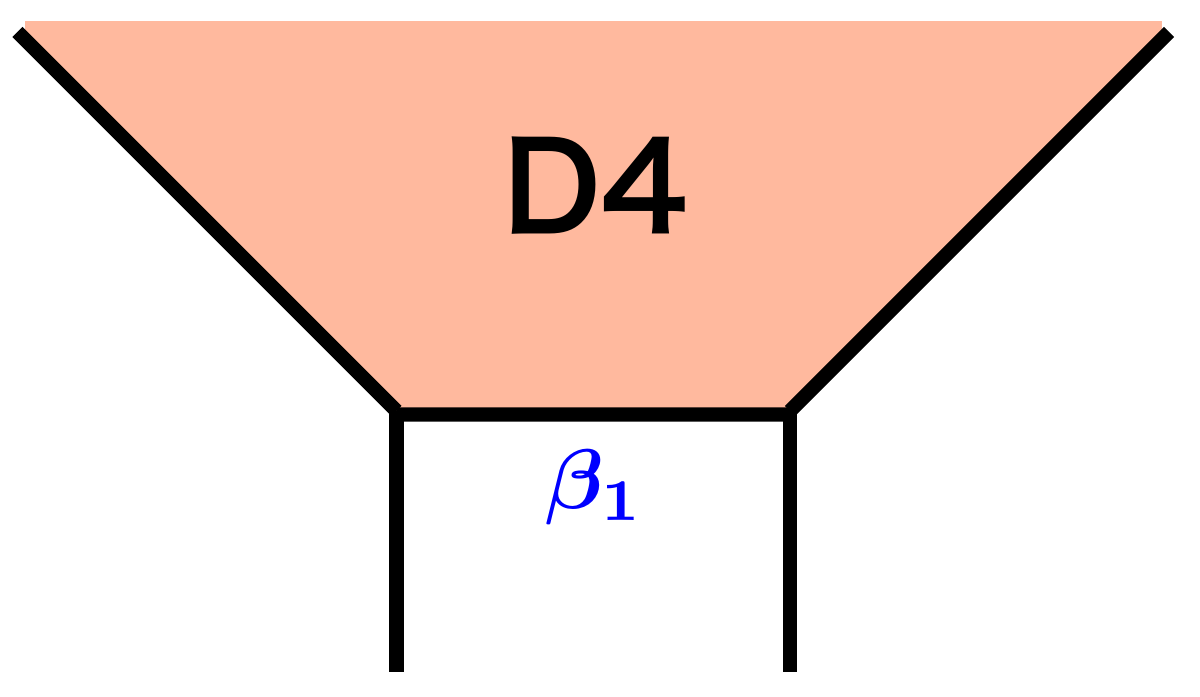}
\caption{The toric webdiagram that has a divisor isomorphic to $A_1$-ALE space.}
\label{fig:toric-A_1}
\end{center}
\end{figure}

The relevant Calabi-Yau three-fold is expressed by figure \ref{fig:toric-A_1}.
We put a single D4-brane on a divisor corresponding to the shaded region in the figure. As seen in the previous section, the D4-D2-D0 bound states in this setup has no wall-crossing phenomena. Therefore, the BPS partition function should always be the same as the instanton partition function on $A_1$-ALE space, which is written in terms of the affine $SU(2)$ character (See equation \eqref{eq:character} and \eqref{eq:Theta_function}):
\begin{eqnarray}
 \chi_{r}^{\widehat{su}(2)_1}(q,Q) &=& \frac{1}{\eta(q)}\sum_{m\in\mathbb{Z}}q^{m^2 + mr + \frac{r^2}{4}}Q^{m+\frac{r}{2}}.\label{eq:instantons_A1}
\end{eqnarray}
Here $q$ denotes the D0-brane charge or equivalently the chemical potential for instantons, while $Q$ is the D2-brane charge or the chemical potential for the magnetic flux on a single blowup cycle in $A_1$-ALE space. The subscript $r = 0, 1$ specifies the representation of the algebra.

It is pointed out in \cite{Vafa:2004qa, Bershadsky:1995qy} that the low energy effective theory on D-branes wrapped on a holomorphic divisor $C_4$ of a Calabi-Yau three-fold is the $d=4,\mathcal{N}=4$ topologically twisted Vafa-Witten theory on $C_4$. The topological twist is necessary to keep the supersymmetry on the curved divisor. By considering the normal bundle to the divisor carefully, we can see that the twist we need is the Vafa-Witten twist. Therefore, the above instanton partition function on a D4-brane should be equivalent to the instanton partition function of the Vafa-Witten theory on $A_1$-ALE space.

\subsection{Dummy cycle and Flop transition}

\begin{figure}
\begin{center}
\includegraphics[width=6cm]{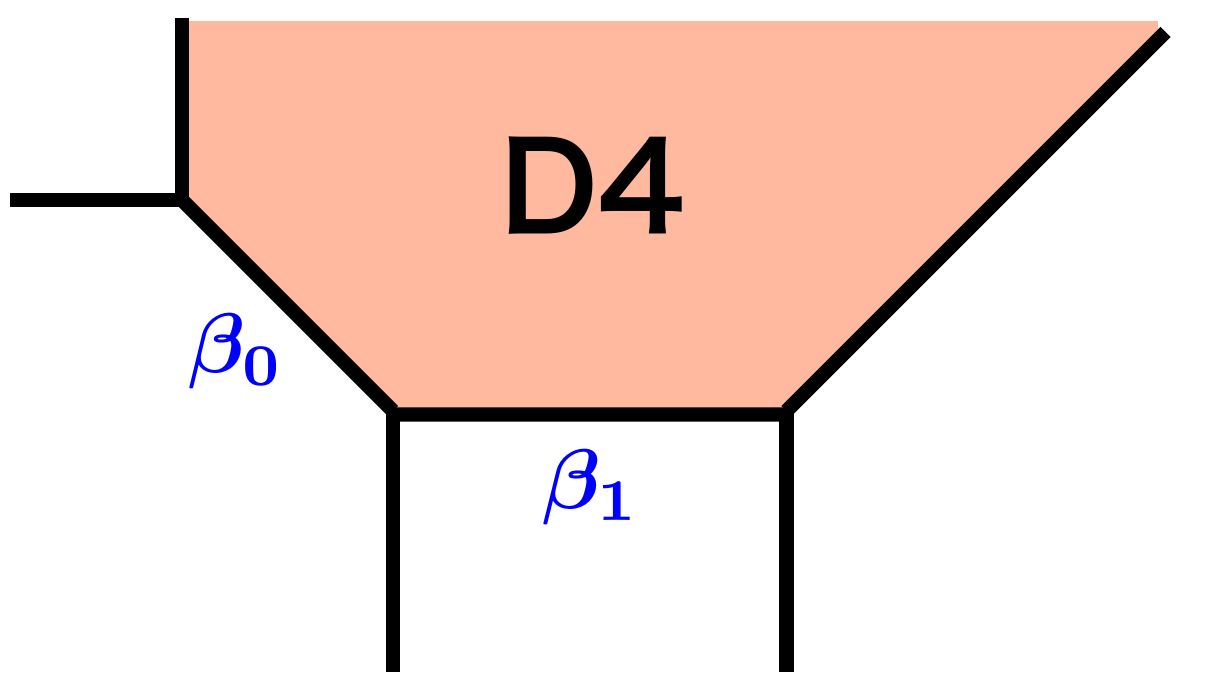}
\caption{A dummy cycle added at an edge of the toric webdiagram.}
\label{fig:dummy_cycle_A1}
\end{center}
\end{figure}

In order to introduce non-vanishing effects of wall-crossings, we here ``add'' an additional dummy two-cycle at an edge of the toric diagram, which drastically changes the situation. The dummy cycle we introduce is denoted by $\beta_0$ in figure \ref{fig:dummy_cycle_A1}.
We set $\beta_0$ so that 
\begin{eqnarray}
 \left\langle \mathcal{D}, \beta_0\right\rangle &=& -1,\label{eq:beta_0}
\end{eqnarray}
which means that in the vicinity of the dummy cycle, the Calabi-Yau $X$ can be regarded as $\mathcal{O}(-1)\oplus\mathcal{O}(-1)\to\mathbb{P}^1$. We also introduce the K\"ahler parameter $z_0$ for the dummy cycle, which modifies the K\"ahler two-form $t$ as
\begin{eqnarray}
 t = z_0\mathcal{P}^0 + z_1\mathcal{P}^1 + \Lambda e^{i\varphi}\mathcal{P}'.
\end{eqnarray}
For large ${\rm Im}\,z_0$, its real and imaginary parts represent the B-field and size of the dummy cycle $\beta_0$.
Here $\mathcal{P}^0$ stands for the harmonic two-form associated with the dummy cycle, that is, we have
\begin{eqnarray}
 \int_X \mathcal{P}^I\wedge \mathcal{\beta}_J = \delta^I_J,
\end{eqnarray}
for $I,J=0,1$. We also impose $\int_X\mathcal{P}'\wedge \beta_0 = 0$. {\em This dummy cycle $\beta_0$ gives us non-vanishing wall-crossing phenomena due to the non-zero intersection product \eqref{eq:beta_0}.} We will study such wall-crossing phenomena in subsection \ref{subsec:walls}.

Note that this modified Calabi-Yau three-fold can be reduced to the original one if we take the non-compact limit ${\rm Im}\,z_0\to +\infty$ for the dummy cycle. Therefore, after taking the limit ${\rm Im}\,z_0\to +\infty$ and omitting the D2-branes wrapped on the dummy cycle $\beta_0$, we should recover the instanton partition functions on $A_1$-ALE space. What is important here is that we should take ${\rm Im}\,z_0$ so that ${\rm Im}\,z_0 \gg {\rm Im}\,z_1$ in order to recover the original Calabi-Yau of figure \ref{fig:toric-A_1}. This means that the large radius limit of the original Calabi-Yau associated with figure \ref{fig:toric-A_1} corresponds to ${\rm Im}\,z_0\to + \infty,\,{\rm Im}\,z_1 \to +\infty$ while keeping $0 < {\rm Im}\,z_1 \ll {\rm Im}\,z_0$. 

\begin{figure}
\begin{center}
\includegraphics[width=6cm]{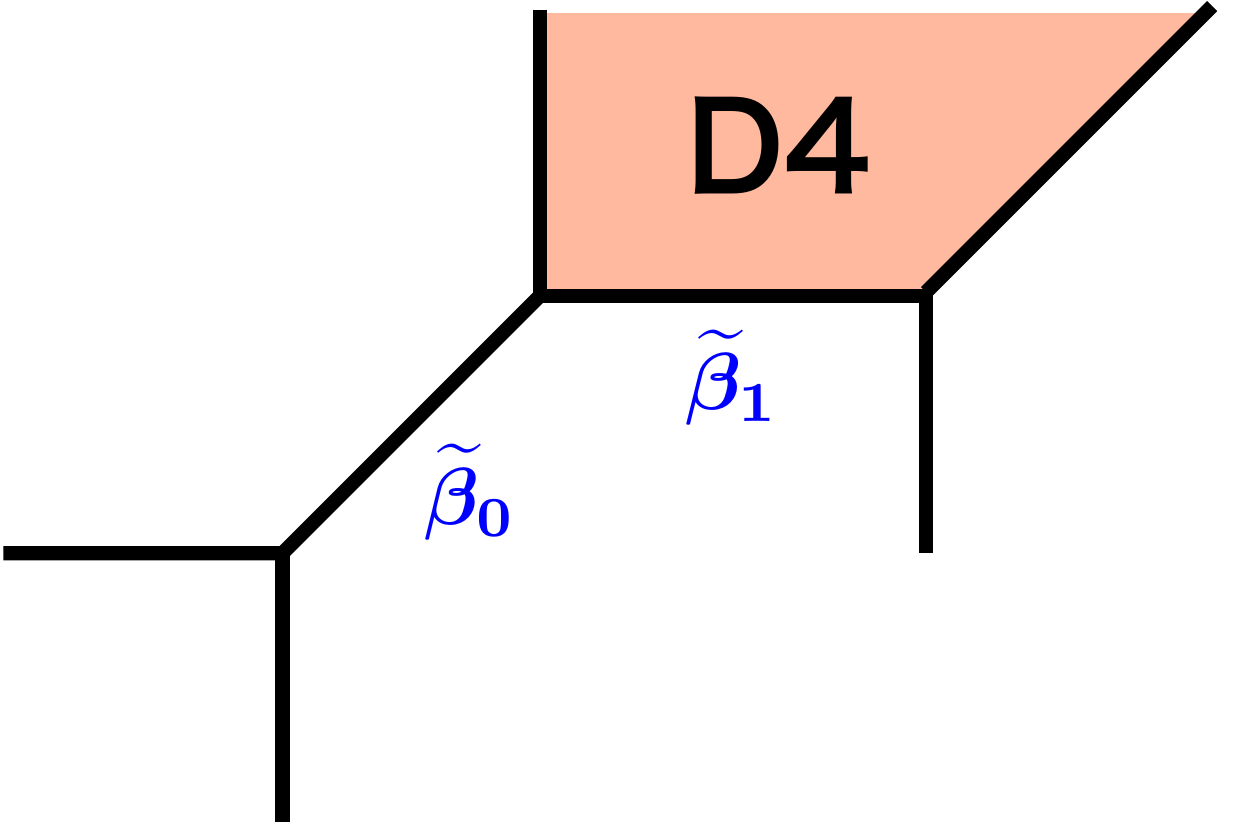}
\caption{The toric webdiagram after the flop transition with respect to the dummy cycle $\beta_0$.}
\label{fig:dummy_flop_A1}
\end{center}
\end{figure}

What is interesting here is that we can consider the negative value of ${\rm Im}\,z_0$. The moduli region ${\rm Im}\,z_0<0$ belongs to a different K\"ahler cone from ${\rm Im}\,z_0>0$, because the dummy cycle vanishes at ${\rm Im}\,z_0 = 0$. In fact, the two moduli regions are connected by the flop transition with respect to the dummy cycle $\beta_0$. The flop transition occurs at ${\rm Im}\,z_0 = 0$ and changes the topology of the Calabi-Yau three-fold as well as that of the divisor wrapped by the D4-brane. After the flop, in the moduli region of ${\rm Im}\,z_0 <0$, the toric diagram of the Calabi-Yau can now be depicted as in figure \ref{fig:dummy_flop_A1}.
We find that now our D4-brane is wrapped on a divisor isomorphic to $\mathcal{O}(-1) \to \mathbb{P}^1$.

As is well-known, flop transitions also change the basis of the two-cycles (For some basic facts about the flop transition, see appendix \ref{app:flop}). For our case, the new basis two-cycles are given by
\begin{eqnarray}
 \widetilde{\beta}_0 = -\beta_0,\qquad \widetilde{\beta}_1 = \beta_1 + \beta_0,
\end{eqnarray}
which implies that 
\begin{eqnarray}
 \langle \mathcal{D},\widetilde{\beta}_0\rangle = 1,\qquad \langle\mathcal{D},\widetilde{\beta}_1\rangle = -1.
\end{eqnarray}
Thus, after the flop transition, the dummy cycle $\widetilde{\beta}_0$ is no longer embedded in the divisor $\mathcal{D}$, and the cycle $\widetilde{\beta_1}$ has a non-vanishing intersection to $\mathcal{D}$. Now, in the vicinity of the cycle $\widetilde{\beta}_1$, our Calabi-Yau is seen as $\mathcal{O}(-1)\oplus\mathcal{O}(-1)\to\mathbb{P}^1$. 

We should introduce the modified basis $\widetilde{\mathcal{P}}^0 = -\mathcal{P}^0+\mathcal{P}^1,\,\widetilde{\mathcal{P}}^1 = \mathcal{P}^1$ for the cycles $\widetilde{\beta}_{0,1}$ so that
\begin{eqnarray}
\int_{X}\widetilde{\mathcal{P}}^I \wedge \widetilde{\beta}_J &=& \delta^I_J,
\end{eqnarray}
for $I,J=0,1$. By using these new basis, the K\"ahler two-form of the Calabi-Yau can be written as
\begin{eqnarray}
 t &=& -z_0\widetilde{\mathcal{P}}^0 + (z_1 + z_0)\widetilde{\mathcal{P}}^1 + \Lambda e^{i\varphi}\mathcal{P}',
\end{eqnarray}
which implies that, in the large radius limit, $\widetilde{z}_0 = -z_0$ and $\widetilde{z}_1 = z_1 + z_0$ stand for the (complexified) areas of the modified two-cycles $\widetilde{\beta}_0$ and $\widetilde{\beta}_1$, respectively.

Now, let us consider what happens if we take a limit of ${\rm Im}\,\widetilde{z}_0 \to +\infty,\, {\rm Im}\,\widetilde{z}_1\to +\infty$. In this limit, the radii of all the compact cycles become infinitely large, and therefore we can again trust the field theory description of our D4-D2-D0 system. All the D2 and D0-brane charges are now realized as magnetic fluxes and instantons on the D4-brane, which is now wrapped on a divisor isomorphic to $\mathcal{O}(-1)\to\mathbb{P}^1$. The corresponding BPS partition function should be equivalent to the instanton partition function of the Vafa-Witten theory on $\mathcal{O}(-1)\to\mathbb{P}^1$, which was evaluated in \cite{Aganagic:2004js}. 

\subsection{Further flop}

\begin{figure}
\begin{center}
\includegraphics[width=4.5cm]{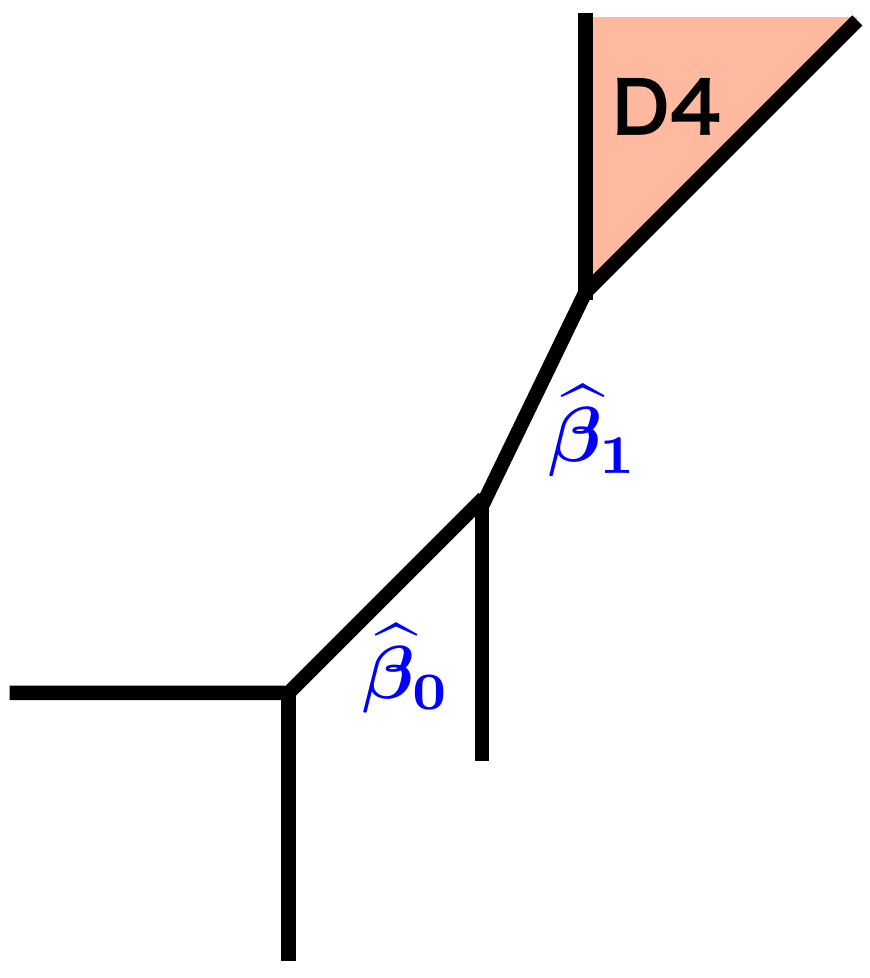}
\caption{The toric webdiagram for the moduli region $0< {\rm Im}\,z_1<-{\rm Im}\,z_0$.}
\label{fig:C2_A1}
\end{center}
\end{figure}

\begin{figure}
\begin{center}
\includegraphics[width=11cm]{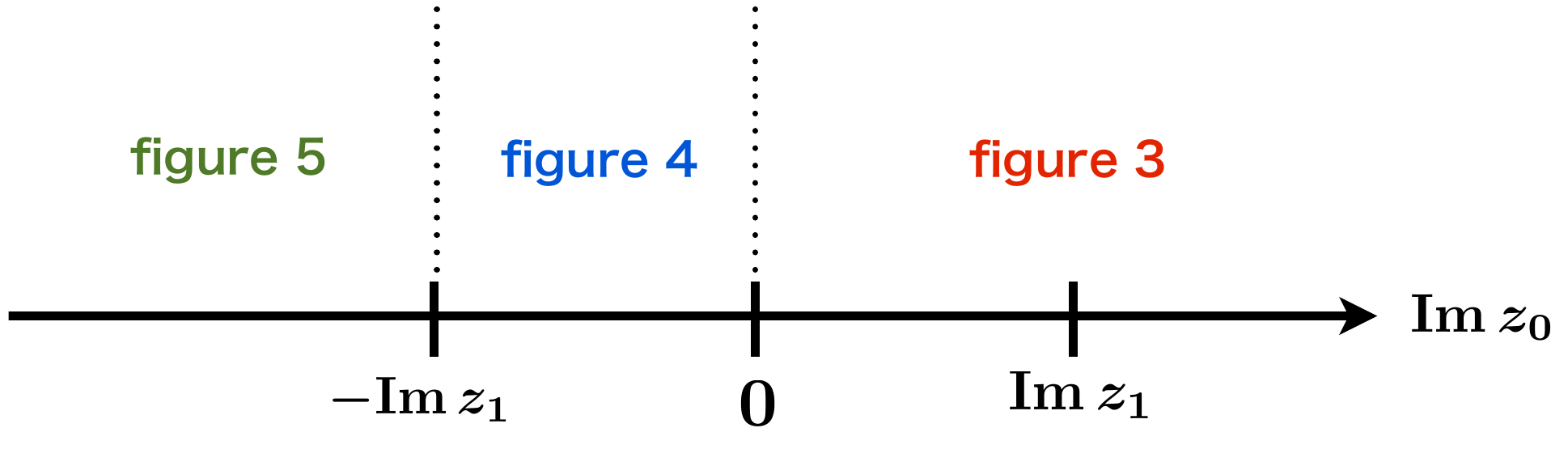}
\caption{Three different regions of the moduli space, which are associated with different K\"ahler cones of the Calabi-Yau three-fold. When ${\rm Im}\,z_0>0$, the toric diagram of the Calabi-Yau is depicted as in figure 3, while the region $-{\rm Im}\,z_1 < {\rm Im}\,z_0 <0$ is associated to figure 4. If ${\rm Im}\,z_0<-{\rm Im}\,z_1$, then the diagram is now depicted as in figure 5.}
\label{fig:three-cones}
\end{center}
\end{figure}

Note here that the above large radii limit ${\rm Im}\,\widetilde{z}_0 \to +\infty,\, {\rm Im}\,\widetilde{z}_1\to +\infty$ implies that ${\rm Im}\,z_0 \to -\infty$ and ${\rm Im}\,z_1 \to +\infty$ while keeping $-{\rm Im}\,z_1\ll {\rm Im}\,z_0 <0$. What happens if we make ${\rm Im}\,z_0$ far smaller than $-{\rm Im}\,z_1$, so that $ {\rm Im}\,z_0 \ll -{\rm Im}\,z_1 < 0$? In that case, ${\rm Im}\,\widetilde{z}_1$ becomes negative, which means that there occurs another flop transition with respect to $\widetilde{\beta}_1$ at ${\rm Im}\,z_0 = -{\rm Im}\,z_1$. Then, the topology of our Calabi-Yau three-fold is again changed, and its toric diagram is now depicted as in figure \ref{fig:C2_A1}.
We find that the divisor wrapped by the D4-brane is now isomorphic to $\mathbb{C}^2$.
In summary we have three different K\"ahler cones associated with three different regions of the moduli space. For fixed ${\rm Im}\,z_1$, the three regions can be depicted as in figure \ref{fig:three-cones}.

According to the rule of the flop transition, the new two-cycles $\widehat{\beta}_0$ and $\widehat{\beta}_1$ in figure \ref{fig:C2_A1} are given by 
\begin{eqnarray}
 \widehat{\beta}_0 = \widetilde{\beta}_0 + \widetilde{\beta}_1 = \beta_1,\qquad \widehat{\beta}_1 = -\widetilde{\beta}_1 = -\beta_0-\beta_1.\label{eq:second_flop_A1}
\end{eqnarray}
Correspondingly, we define $\widehat{\mathcal{P}}^0 = -\mathcal{P}^0 +  \mathcal{P}^1$ and $\widehat{\mathcal{P}}^1 = -\mathcal{P}^0$ so that
\begin{eqnarray}
 \int_X \widehat{\mathcal{P}}^I \wedge \widehat{\beta}_J &=& \delta^I_J
\end{eqnarray}
for $I, J = 0,1$. By using these new basis harmonic two-forms, we can expand the K\"ahler two-form as
\begin{eqnarray}
 t &=& \widehat{z}_0\widehat{\mathcal{P}}^0 + \widehat{z}_1\widehat{\mathcal{P}}^1 + \Lambda e^{i\varphi}\mathcal{P}',
\end{eqnarray}
where we defined $\widehat{z}_0 = z_1$ and $\widehat{z}_1 = -z_0-z_1$. Thus, ${\rm Im}\,z_1$ and ${\rm Im}\,(-z_0-z_1)$ stand for the radii of the new two-cycles $\widehat{\beta}_0$ and $\widehat{\beta}_1$ respectively, if both are large enough. Note that both of the quantities are always positive in the moduli region ${\rm Im}\,z_0 < -{\rm Im}\,z_1 < 0$.

Now, let us consider the large radii limit of the Calabi-Yau of figure \ref{fig:C2_A1}. Since the radii of the two-cycles are now ${\rm Im}\,\widehat{z}_0={\rm Im}\,z_1$ and ${\rm Im}\,\widehat{z}_1={\rm Im}\,(-z_0-z_1)$, the large radii limit implies ${\rm Im}\,z_0 \to -\infty,\,{\rm Im}\,z_1\to +\infty$ while keeping ${\rm Im}\,z_0 \ll -{\rm Im}\,z_1$. Recall that the divisor wrapped by the D4-brane is now isomorphic to $\mathbb{C}^2$. Thus, in this limit, the BPS partition function of our D4-D2-D0 states should be equal to the instanton partition function on $\mathbb{C}^2$. It is well-known that the instanton partition function of the Vafa-Witten theory on $\mathbb{C}^2$ is given by
\begin{eqnarray}
 \mathcal{Z}_{\mathbb{C}^2}(q,Q) &=& \prod_{n=1}^\infty\frac{1}{1-q^n} = \frac{1}{\phi(q)},
\label{eq:C2_A1}
\end{eqnarray}
where $\phi(q) = \prod_{n=1}^\infty(1-q^n)$ is the Euler function.

\subsection{Walls of marginal stability}
\label{subsec:walls}

Now, we have two special large radii limits: (i)$\;{\rm Im}\,z_0\to -\infty,\,{\rm Im}\,z_1\to +\infty$ with ${\rm Im}\,z_0 \ll -{\rm Im}\,z_1$; and (ii)$\; {\rm Im\,}z_0 \to +\infty,\,{\rm Im}\,z_1 \to +\infty$ with ${\rm Im}\,z_1 \ll {\rm Im}\,z_0$. The former leads to \eqref{eq:C2_A1}, while the latter gives us the affine $SU(2)$ character \eqref{eq:instantons_A1} after neglecting the effects of the dummy cycle $\beta_0$. These two large radii limits can be connected by changing the moduli parameters, which gives rise to the wall-crossing phenomena of D4-D2-D0 states. By using the wall-crossing formula, we can calculate the change of the BPS partition function along the wall-crossings between (i) and (ii). If the wall-crossing formula correctly works, such changes in the BPS partition function {\em should interpolate two different instanton partition functions \eqref{eq:C2_A1} and \eqref{eq:instantons_A1}.} In the rest of this section, we will explicitly verify this, and show that the affine $SU(2)$-character \eqref{eq:instantons_A1} is obtained from the instanton partition function on $\mathbb{C}^2$ \eqref{eq:C2_A1}, by considering the wall-crossing of D4-D2-D0 states.

Recall that the relevant wall-crossings are associated with separations of D2-D0 fragments from our D4-D2-D0 states. The wall-crossing formula we should use is \eqref{eq:wall-crossing_formula}. To use the formula, we should learn the BPS index $\Omega(\Gamma_2)$ for
\begin{eqnarray}
\Gamma_2 = m^0\beta_0 + m^1\beta_1 - ndV,
\end{eqnarray}
where $m^i\in \mathbb{Z}$ denote the D2-brane charges for the two-cycles $\beta_i$. The BPS index of the D2-D0 states on a Calabi-Yau three-fold can be read off from the Gopakumar-Vafa invariants \cite{Gopakumar:1998ii, Gopakumar:1998jq}. For our toric Calabi-Yau three-fold of figure \ref{fig:dummy_cycle_A1}, the only non-vanishing Gopakumar-Vafa invariants $N^{\beta}_r$ are
\begin{eqnarray}
 N^{\beta}_0 &=& 1\quad {\rm for}\quad \beta = \beta_0 \quad {\rm and}\quad \beta = \beta_0+ \beta_1,
\nonumber \\
N^{\beta}_0 &=& -1\quad {\rm for}\quad \beta = \beta_1.\label{eq:GV-A1}
\end{eqnarray}
Note that all the higher genus contributions vanish.
For the derivation of this result, see appendix \ref{app:Gopakumar-Vafa}.
The Gopakumar-Vafa invariants physically represent the BPS indices of M2-branes wrapped on two-cycles $\beta$ of the Calabi-Yau three-fold, which can be seen as D2-D0 indices after dimensionally reducing the M-theory circle transverse to the Calabi-Yau. Therefore, the Gopakumar-Vafa invariants \eqref{eq:GV-A1} imply that the only non-vanishing D2-D0 indices are
\begin{eqnarray}
 \Omega(\pm \beta_0 - ndV) = \Omega(\pm \beta_0 \pm \beta_1-ndV) = 1,\qquad \Omega(\pm \beta_1-ndV) = -1,\label{eq:D2-D0_A1}
\end{eqnarray}
where the D0-brane charge $n$ comes from the KK-momentum along the M-theory circle, and the charge conjugation is taken into account.\footnote{In addition to these D2-D0 indices, there are still non-vanishing indices for pure D0-branes, but they are irrelevant because pure D0-branes have vanishing charge intersection products with our D4-D2-D0 states. If we introduce D6-branes, then we have to take into account decays into pure D0-branes.}

Here, let us denote by $W^{m^0,m^1}_n$ the wall of marginal stability associated with a decay channel
\begin{eqnarray}
\Gamma = \mathcal{D}+k^I\beta_I - ldV \quad \to\quad (\Gamma_1 = \Gamma - \Gamma_2) \quad +\quad (\Gamma_2 = m^I\beta_I - ndV).
\end{eqnarray}
The equation \eqref{eq:D2-D0_A1} implies that the only candidates of walls that might give rise to any changes in the BPS partition function are $W^{\pm1,0}_n,\,W^{\pm1,\pm1}_n$ and $W^{0,\pm1}_n$. Other types of the walls give no change in the BPS partition function due to their vanishing indices. Furthermore, it follows that the walls of $W^{0,\pm1}_n$ also give no change in the BPS partition function, because the charge $\Gamma_2 = \pm\beta_1+ndV$ has the vanishing intersection product with $\Gamma$. Hence, we find that it is sufficient to consider only the walls of $W^{\pm1,0}_n, W^{\pm1,\pm1}_n$.

We now identify the locations of the relevant walls $W^{\pm1,0}_n,\,W^{\pm1,\pm1}_n$ in the moduli space. Recall that the walls of marginal stability are subspace in the moduli space where $\arg Z(\Gamma_1) = \arg Z(\Gamma_2)$ is satisfied. In our case, the two central charges are evaluated as
\begin{eqnarray}
 Z(\Gamma_1) \sim -\frac{c_4}{2}\Lambda^2 e^{2i\varphi},\qquad Z(\Gamma_2) = m^0z_0 + m^1z_1 + n,
\end{eqnarray} 
where $Z(\Gamma_1)$ is dominated by the D4-brane contribution in the local limit $\Lambda\to +\infty$. We fix $\varphi$ and regard the space of $(z_0,z_1)$ as our moduli space. If we move the moduli $(z_0,z_1)$ and $\arg Z(\Gamma_1) = \arg Z(\Gamma_2)$ is satisfied for some $(m^0,m^1)\in\{(\pm1,0),(\pm1,\pm1)\}$ and $n\in \mathbb{Z}$, then the BPS index of our D4-D2-D0 states might be changed discontinuously. The locations of the walls of marginal stability $W^{m^0,m^1}_n$ are identified as
\begin{eqnarray}
 2\varphi &=& \arg\left(-m^0z_0 - m^1z_1 -n\right).
\end{eqnarray}
From this, we can draw the relevant walls of marginal stability $W^{\pm1,0}_n$ and $W^{\pm1,\pm1}_n$ in the moduli space. 

\begin{figure}
\begin{center}
\includegraphics[width=14.5cm]{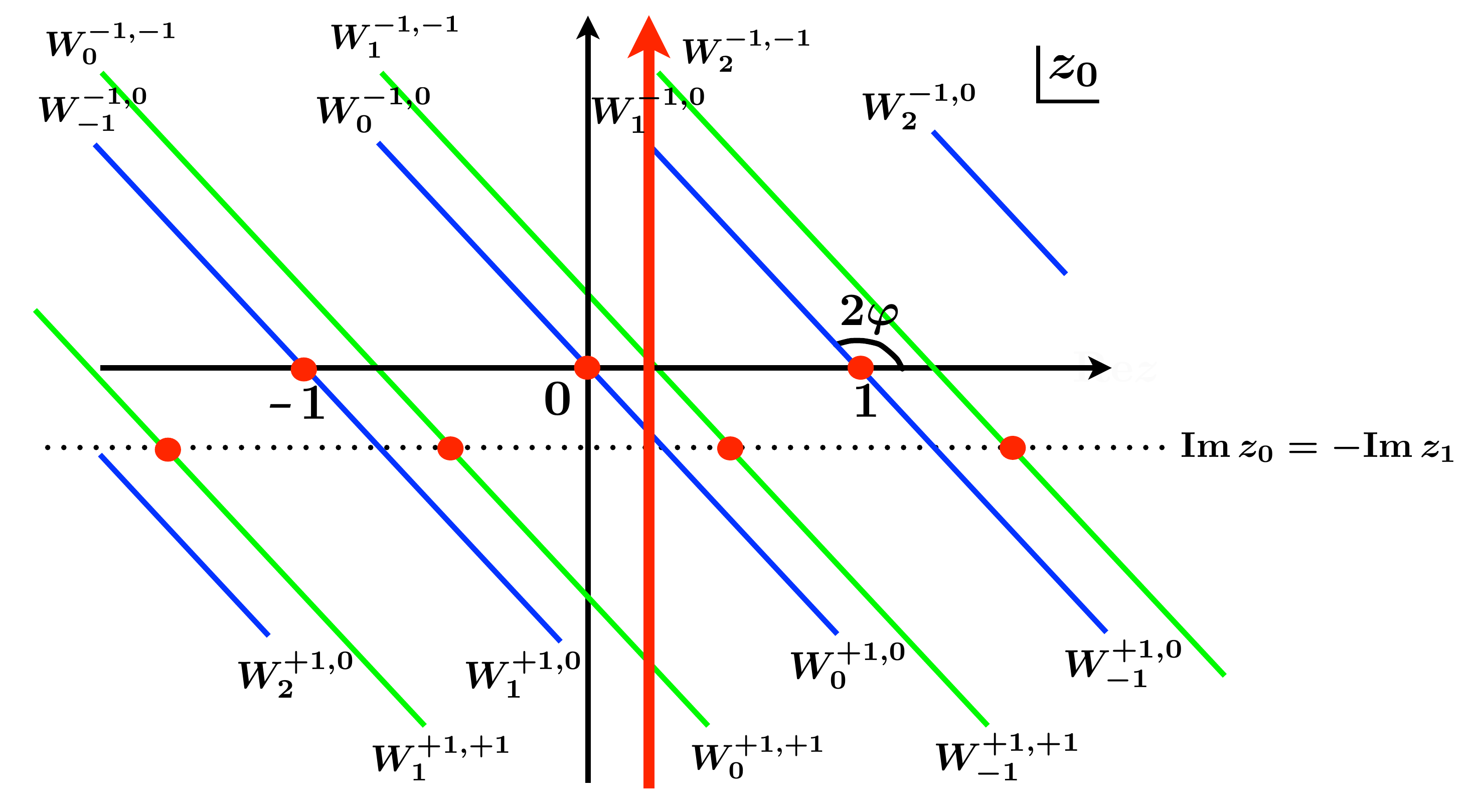}
\caption{The walls of marginal stability in $z_0$-plane with $z_1=1/2$ fixed. All the walls are straight lines whose slope is $2\varphi$. Note that we have fixed $\varphi$ so that $\pi/4 < \varphi < \pi/2$. The red dots denote singularities where some D2-D0 state becomes massless.}
\label{fig:walls_A1}
\end{center}
\end{figure}

Let us draw the walls in $z_0$-plane with $z_1$ fixed, as in figure \ref{fig:walls_A1}. 
All the walls are straight lines whose slope is $2\varphi$. The blue walls are associated with $W^{\pm1,0}_n$ while the green ones correspond to $W^{\pm1,\pm1}_n$. The red dots stand for the singularities in the moduli space where some D2-D0 state becomes massless. In this paper, we always tune the moduli so that they do not cross the massless singularities.
When we change the value of the modulus $z_1$, the green walls associated with $W^{\pm1,\pm1}_N$ move in $z_0$-plane, while the locations of the blue walls for $W^{\pm1,0}_N$ remain unchanged. In particular, changing ${\rm Im}\,z_1$ leads to moving the green walls {\em vertically.} Below, we always fix ${\rm Re}\,z_1$ so that $0 < {\rm Re}\,z_1<1$, and sometimes move ${\rm Im}\,z_1$.\footnote{The shift of ${\rm Re}\,z_1$ by some integer gives rise to monodromy, but we will not discuss it in this paper.}

As mentioned at the beginning of this subsection, our aim is to investigate the wall-crossings occurring when the moduli move from (i) ${\rm Im}\,z_0 \to - \infty,\,{\rm Im}\,z_1\to +\infty$ with ${\rm Im}\,z_0\ll-{\rm Im}\,z_1<0$ to (ii) ${\rm Im}\,z_0 \to +\infty,\,{\rm Im}\,z_1 \to +\infty$ with $0<{\rm Im}\,z_1 \ll {\rm Im}\,z_0$. Between (i) and (ii), we move $z_0$ along the red arrow in figure \ref{fig:walls_A1}. Along the red arrow, the moduli cross all the walls of\footnote{Recall that we have fixed $\varphi$ so that $\pi/4 < \varphi < \pi/2$.}
\begin{eqnarray}
&& W^{+1,0}_{n},\;W^{+1,+1}_{n} \quad {\rm for} \quad n=0,1,2,\cdots,\label{eq:walls1_A1}
\\
&& W^{-1,0}_{n},\;W^{-1,-1}_{n} \quad {\rm for} \quad n=1,2,3,\cdots.\label{eq:walls2_A1}
\end{eqnarray}

One might think that the walls $W^{\pm1,\pm1}_n$ are not crossed because in the limit of ${\rm Im}\,z_1\to+\infty$ the green walls move out to the infinity of the lower half $z_0$-plane. However, due to the condition ${\rm Im}\,z_0\ll-{\rm Im}\,z_1<0$ in (i), ${\rm Im}\,z_0$ is far smaller than $-{\rm Im}\,z_1$ at the bottom of the red arrow, and therefore all the walls of $W^{-1,-1}_{n\geq 1}$ and $W^{+1,+1}_{n\geq 0}$ must be crossed between (i) and (ii). To be more precise, we first fix the modulus ${\rm Im}\,z_1 \gg 1$, and move the other modulus ${\rm Im}\,z_0$ from ${\rm Im}\,z_0 = -\infty (\ll -{\rm Im}\,z_1)$ to ${\rm Im}\,z_0 = + \infty$. Then all the walls of \eqref{eq:walls1_A1} and \eqref{eq:walls2_A1} are crossed. Recall that the condition ${\rm Im}\,z_0 \ll -{\rm Im}\,z_1$ at the bottom of the arrow corresponds to the large radius limit of $\widehat{\beta}_1$, and needs to be satisfied in order to reach the field theory limit in (i).

\subsection{Affine $SU(2)$ character}

We now come to the main result of this section. We here evaluate the discrete changes in the BPS partition function that occur when the moduli move from (i) to (ii), and show that they properly reproduce the affine $SU(2)$ character living in (ii). As shown in the previous subsection, the moduli cross the walls of \eqref{eq:walls1_A1} and \eqref{eq:walls2_A1} between (i) and (ii). The BPS partition function jumps at each of the walls, following the wall-crossing formula \eqref{eq:wall-crossing_formula}.

Recall that the form of \eqref{eq:wall-crossing_formula} is valid only when the ordering of the central charge phases are reversed from $\arg Z(\Gamma_2)<\arg Z(\Gamma_1)$ to $\arg Z(\Gamma_2)>\arg Z(\Gamma_1)$ at the wall-crossing. In our case, $Z(\Gamma_1)\sim -\frac{c_4}{2}\Lambda^2e^{2i\varphi}$ and $Z(\Gamma_2) = m^0z_0 + m^1z_1 + n$ for the wall $W^{m^0,m^1}_n$. Thus, we find that \eqref{eq:wall-crossing_formula} is applicable if $m^0{\rm Im}\,z_0$ {\em increases} along the red arrow in figure \ref{fig:walls_A1}. If $m^0{\rm Im}\,z_0$ {\em decreases} along the arrow, we should reverse the sign of the exponent in \eqref{eq:wall-crossing_formula}. The difference between these two corresponds to whether the BPS bound state disappear or newly appear in the spectrum.

We now move the moduli along the red axis in figure \ref{fig:walls_A1}. When the moduli cross the walls of \eqref{eq:walls1_A1}, the quantity $m^0\,{\rm Im}\,z_0$ increases along the red arrow. Therefore, they give rise to a multiplication of
\begin{eqnarray}
 \prod_{n=0}^\infty \left(1-q^nQ_0\right)\left(1-q^nQ_0Q_1\right)\label{eq:wall-crossing1_A1}
\end{eqnarray}
to the BPS partition function, where we used \eqref{eq:D2-D0_A1} and the fact that $\langle\Gamma_2,\Gamma\rangle=1$.\footnote{Note here that now $\Gamma_2$ itself is primitive, and therefore $\Gamma_2' = \Gamma_2$ in \eqref{eq:wall-crossing_formula}.} Here $Q_0, Q_1$ are chemical potentials for D2-branes wrapped on $\beta_0,\beta_1$, respectively, and $q$ denotes the D0-brane chemical potential. On the other hand, when the moduli cross the walls of \eqref{eq:walls2_A1}, $m^0\,{\rm Im}\,z_0$ decreases along the red arrow. So we should reverse the sign of the exponent in \eqref{eq:wall-crossing_formula} when using the wall-crossing formula. However, simultaneously, we now have an opposite sign of the intersection product $\langle\Gamma_2,\Gamma\rangle = -1$, which cancels the reversed sign of the exponent. Thus, crossing the walls \eqref{eq:walls2_A1} gives the following multiplication to the partition function:
\begin{eqnarray}
 \prod_{n=1}^\infty\left(1-q^nQ_0^{-1}\right)\left(1-q^nQ_0^{-1}Q_1^{-1}\right).
\label{eq:wall-crossing2_A1}
\end{eqnarray}
Hence, in total, we have two multiplicative contributions \eqref{eq:wall-crossing1_A1} and \eqref{eq:wall-crossing2_A1} along the red arrow in figure \ref{fig:walls_A1}. Now, let $\mathcal{Z}_{-\infty}(q,Q)$ be the BPS partition function in the limit of (i), and $\mathcal{Z}_{+\infty}(q,Q)$ be that in the limit of (ii). The above argument implies that these two partition functions are connected by the relation
\begin{eqnarray}
 \mathcal{Z}_{+\infty}(q,Q) &=& \mathcal{Z}_{-\infty}(q,Q) \prod_{n=0}^\infty \left(1-q^nQ_0\right)\left(1-q^nQ_0Q_1\right) \prod_{m=1}^\infty\left(1-q^mQ_0^{-1}\right)\left(1-q^mQ_0^{-1}Q_1^{-1}\right).
\nonumber \\ \label{eq:relation_A1}
\end{eqnarray}

Recall here that $\mathcal{Z}_{-\infty}$ is equal to \eqref{eq:C2_A1}, and $\mathcal{Z}_{+\infty}$ should be written by affine $SU(2)$ character \eqref{eq:instantons_A1} after neglecting the effects of dummy cycle $\beta_0$. Below, we will explicitly verify this. We start from
\begin{eqnarray}
 \mathcal{Z}_{-\infty}(q,Q) &=& \prod_{n=1}^\infty \frac{1}{1-q^n} = \frac{1}{\phi(q)},
\end{eqnarray}
and substitute it into \eqref{eq:relation_A1} to obtain
\begin{eqnarray}
 \mathcal{Z}_{+\infty}(q,Q) &=& \frac{1}{\phi(q)^3}\sum_{n_0,n_1\in\mathbb{Z}}q^{\frac{1}{2}n_0(n_0-1) + \frac{1}{2}n_1(n_1-1)}(-Q_0)^{n_0}(-Q_0Q_1)^{n_1}.\label{eq:pre-result_A1}
\end{eqnarray}
Here we used an identity
\begin{eqnarray}
 \prod_{l=1}^\infty(1-q^l)\prod_{m=0}^\infty(1-q^mQ)\prod_{n=1}^\infty(1-q^nQ^{-1}) &=& \sum_{n\in\mathbb{Z}}q^{\frac{1}{2}n(n-1)}(-Q)^n,\label{eq:identity}
\end{eqnarray}
which follows from the Jacobi's triple product identity. Now, we extract the $Q_0$-independent terms from \eqref{eq:pre-result_A1} in order to neglect the contributions from D2-branes wrapped on the dummy cycle $\beta_0$. The result is
\begin{eqnarray}
\left.\mathcal{Z}_{+\infty}(q,Q)\right|_{Q_0-{\rm independent}} = \frac{1}{\phi(q)^3}\sum_{n\in\mathbb{Z}}q^{n^2}Q_1^{n} = \frac{q^{1/8}}{\eta(q)^2}\,\chi^{\widehat{su}(2)_1}_0(q,Q_1).
\end{eqnarray}
This is correctly proportional to the level one affine $SU(2)$ character $\chi^{\widehat{su}(2)_1}_r$ with $r=0$. The prefactor is related to the D4-D0 bound states and independent of the D2-brane chemical potential $Q_1$. The $Q_1$-dependent terms are perfectly given by the character of $\widehat{su}(2)$. This result implies that the wall-crossing formula \eqref{eq:wall-crossing_formula} knows about the affine $SU(2)$ algebra acting on the cohomology of the instanton moduli space of $A_1$-ALE space.

Finally, let us briefly comment on the parameter $r$. The parameter $r$ labels the representation of $\widehat{su}(2)$, which should be related to the boundary condition at infinity of the divisor wrapped by the D4-brane. In \cite{Dijkgraaf:2007sw}, it was argued that the parameter $r$ depends on the boundary condition at infinity. It would be interesting to perform further study on the boundary condition.

\section{Affine $SU(N)$ algebra from wall-crossings}
\label{sec:affine_SU(N)}

\subsection{Dummy cycle and Flop transitions}

In this section, we generalize the previous argument to the relation between affine $SU(N)$ algebra and $A_{N-1}$-ALE space. The toric diagram of the relevant toric Calabi-Yau three-fold is depicted in figure \ref{fig:toric-A_{N-1}}. We put a single D4-brane on a divisor isomorphic to $A_{N-1}$-ALE space, whose projection onto the toric base is a shaded region of figure \ref{fig:toric-A_{N-1}}.

\begin{figure}
\begin{center}
\includegraphics[width=6cm]{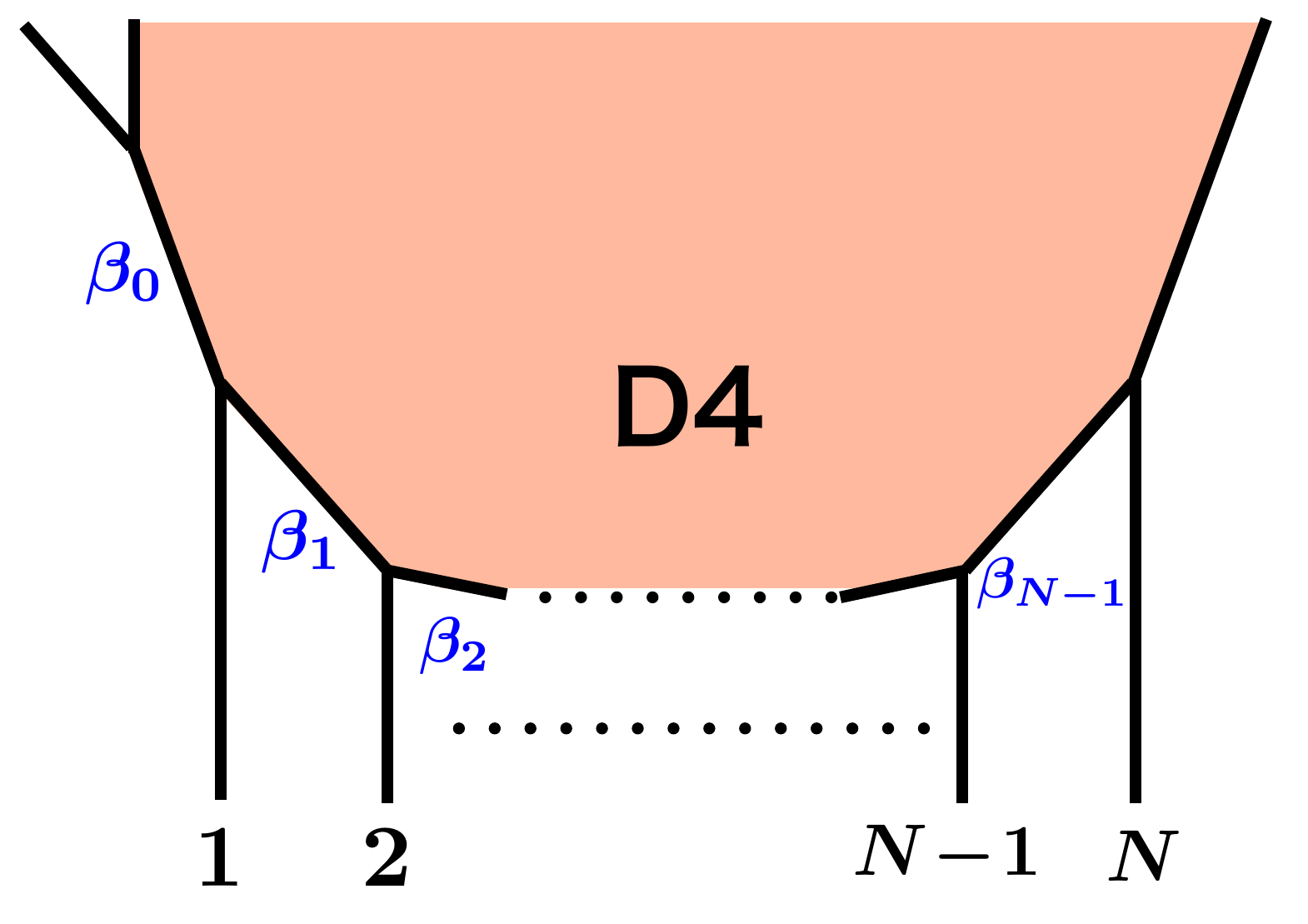}
\caption{The toric diagram after adding a dummy cycle at an edge of the original diagram. The dummy cycle is denoted by $\beta_0$.}
\label{fig:dummy_cycle}
\end{center}
\end{figure}

As shown in section \ref{sec:wall-crossing}, our D4-D2-D0 states have no wall-crossing with respect to the moduli $z^i$ for $i=1,2,\cdots, N-1$. So, as in the case of $A_1$-ALE space, we add a ``dummy'' two-cycle at an edge of the toric diagram. Let the dummy cycle be labeled by $\beta_0$ as in figure \ref{fig:dummy_cycle}, where we impose
\begin{eqnarray}
 \left\langle \mathcal{D}, \beta_0\right\rangle &=& -1,
\end{eqnarray}
which means that in the vicinity of the dummy cycle in the toric base, the Calabi-Yau $X$ can be regarded as $\mathcal{O}(-1)\oplus\mathcal{O}(-1)\to\mathbb{P}^1$. Including the K\"ahler parameter $z_0$ for the dummy cycle, the K\"ahler two-form is modified as
\begin{eqnarray}
 t = z_0\mathcal{P}^0 + \sum_{i=1}^{N-1} z_i\mathcal{P}^i + \Lambda e^{i\varphi}\mathcal{P}',
\end{eqnarray}
where $\mathcal{P}^0$ stands for the harmonic two-form associated to the dummy cycle, that is, it follows that
\begin{eqnarray}
 \int_X \mathcal{P}^I\wedge \mathcal{\beta}_J &=& \delta^I_J
\end{eqnarray}
for $I,J=0,1,2,\cdots, N-1$. This modified Calabi-Yau three-fold is reduced to the original one if we take the non-compact limit ${\rm Im}\,z_0\to +\infty$ for the dummy cycle. In particular, after taking the limit ${\rm Im}\,z_0\to +\infty$ and neglecting the effects of the dummy cycle, we should recover the D4-D2-D0 states on $A_{N-1}$-ALE space.

\begin{figure}
\begin{center}
\includegraphics[width=8cm]{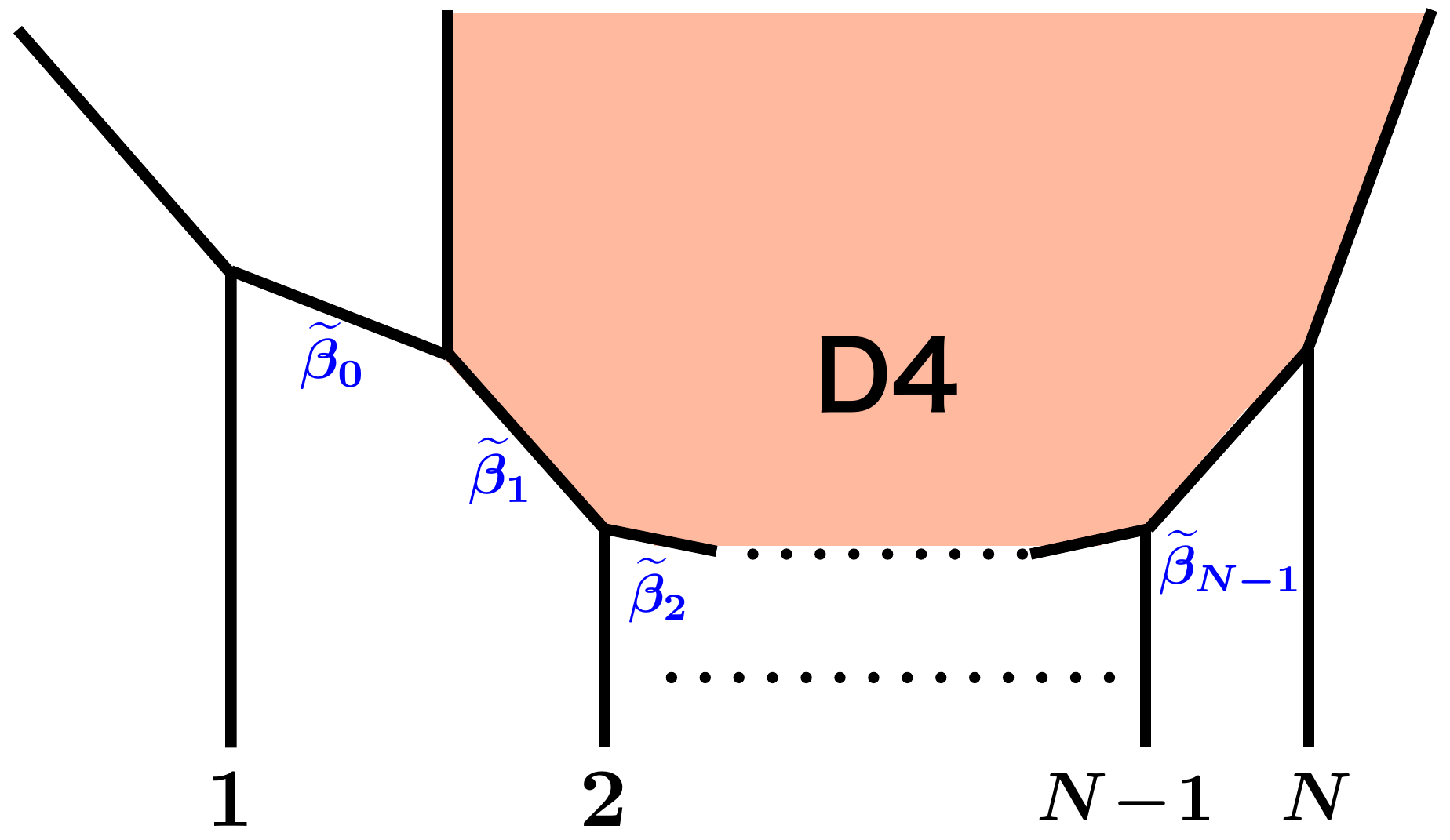}
\caption{The toric webdiagram after the flop transition with respect to the dummy cycle $\beta_0$.}
\label{fig:dummy_flop}
\end{center}
\end{figure}

As in the case of $A_1$-ALE space, two moduli regions ${\rm Im}\,z_0>0$ and ${\rm Im}\,z_0 < 0$ are connected by the flop transition with respect to the dummy cycle. If we decrease ${\rm Im}\,z_0$ through ${\rm Im}\,z_0=0$, our Calabi-Yau becomes now described by the diagram in figure \ref{fig:dummy_flop}.
Correspondingly, the basis of the second homologies should be modified as
$
\widetilde{\beta}_0 = -\beta_0,\; \widetilde{\beta}_1 = \beta_1 + \beta_0,\; \widetilde{\beta}_i = \beta_i$ for $i=2,3,\cdots,N-1$, which implies that 
\begin{eqnarray}
 \langle \mathcal{D},\widetilde{\beta}_0\rangle = 1,\qquad \langle\mathcal{D},\widetilde{\beta}_1\rangle = -1,\qquad \langle\mathcal{D},\widetilde{\beta}_i\rangle = 0 \quad {\rm for} \quad i=2,3,\cdots, N-1.
\end{eqnarray}
The modified basis two-forms are defined by 
$
\widetilde{\mathcal{P}}^0 = -\mathcal{P}^0+\mathcal{P}^1,\;\widetilde{\mathcal{P}}^1 = \mathcal{P}^1,\; \widetilde{\mathcal{P}}^{i} = \mathcal{P}^i$ for $i = 2,3,\cdots,N-1$, so that
\begin{eqnarray}
\int_{X}\widetilde{\mathcal{P}}^I \wedge \widetilde{\beta}_J &=& \delta^I_J,
\end{eqnarray}
for $I,J=0,1,\cdots,N-1$. By using this new basis, the K\"ahler two-form of the Calabi-Yau can be written as
\begin{eqnarray}
 t &=& \widetilde{z}_0\widetilde{\mathcal{P}}^0 + \widetilde{z}_1\widetilde{\mathcal{P}}^1 + \sum_{i=2}^{N-1}\widetilde{z}_i\widetilde{\mathcal{P}}^i + \Lambda e^{i\varphi}\mathcal{P}',
\end{eqnarray}
where $\widetilde{z}_0 = -z_0,\;\widetilde{z}_1 = z_1 + z_0$ and $\widetilde{z}_i = z_i$ for $i=2,3,\cdots,N-1$. When ${\rm Im}\,\widetilde{z}_I$ is large, the real and imaginary parts of $\widetilde{z}_I$ represent the size and B-field of the new two-cycle $\widetilde{\beta}_I$, respectively.

As we keep decreasing ${\rm Im}\,z_0$, we have another flop transition with respect to $\widetilde{\beta}_1$ at ${\rm Im}\,z_0 = -{\rm Im}\,z_1$, where ${\rm Im}\,\widetilde{z}_1$ becomes negative and the topology of our Calabi-Yau again changes.
 As in equation \eqref{eq:second_flop_A1}, such a flop leads us to define further modified basis two-cycles $\widehat{\beta}_I$ as
\begin{eqnarray}
\widehat{\beta}_0 &=& \widetilde{\beta}_0 + \widetilde{\beta}_1 = \beta_1,\qquad \widehat{\beta}_1 = - \widetilde{\beta}_1 = -\beta_0 -\beta_1,
\nonumber \\[2mm]
 \widehat{\beta}_2  &=& \widetilde{\beta}_1 + \widetilde{\beta}_2 =  \beta_0 +\beta_1+\beta_2,
\nonumber \\[2mm]
\widehat{\beta}_i &=& \beta_i \quad {\rm for}\quad i=3,4,\cdots,N-1,
\end{eqnarray}
which implies new basis of the K\"ahler parameters
\begin{eqnarray}
&&\widehat{z}_0 = z_1,\qquad \widehat{z}_1 = -z_0-z_1,\qquad \widehat{z}_2 = z_0+z_1+z_2,
\nonumber \\[2mm]
&&\widehat{z}_i = z_i \quad {\rm for}\quad i=3,4,\cdots, N-1.
\end{eqnarray}
From this, we find that there is yet another flop transition at ${\rm Im}(z_0+z_1+z_2) = 0$.
 In general, there are $N$ different flop transitions at ${\rm Im}\,z_0 = -\sum_{i=1}^{k}{\rm Im}\,z_i$ for $k=0,1,2,\cdots,N-1$. For each of these flops, we have a topology-change of our Calabi-Yau. Therefore, {\em we have $N+1$ topologically different resolutions of the Calabi-Yau three-fold.} The moduli space is divided into $N+1$ regions, which correspond to different K\"ahler cones of the Calabi-Yau and connected by flop transitions. In $z_0$-plane with $z_1,\cdots,z_{N-1}$ fixed, such $N-1$ regions are given by
\begin{eqnarray}
R_0 &=& \left\{z_0\;\left|\; {\rm Im}\,z_0>0\right.\right\},\qquad
R_{1} = \left\{z_0\;\left|\; -{\rm Im}\,z_1< {\rm Im}\,z_0 <0\right.\right\}.\label{eq:cones1_AN}
\\
R_k &=& \left\{z_0\;\left|\;  -\sum_{i=1}^k {\rm Im}\,z_i \;<\; {\rm Im}\,z_0 \;<\; -\sum_{i=1}^{k-1}{\rm Im}\,z_i\;<\;0\right.\right\},\label{eq:cones2_AN}
\\
R_{N} &=& \left\{z_0\;\left|\; {\rm Im}\,z_0 \;<\; -\sum_{i=1}^{N-1}{\rm Im}\,z_i\;<\;0\right.\right\}.\label{eq:cones3_AN}
\end{eqnarray}

\begin{figure}
\begin{center}
\includegraphics[width=7cm]{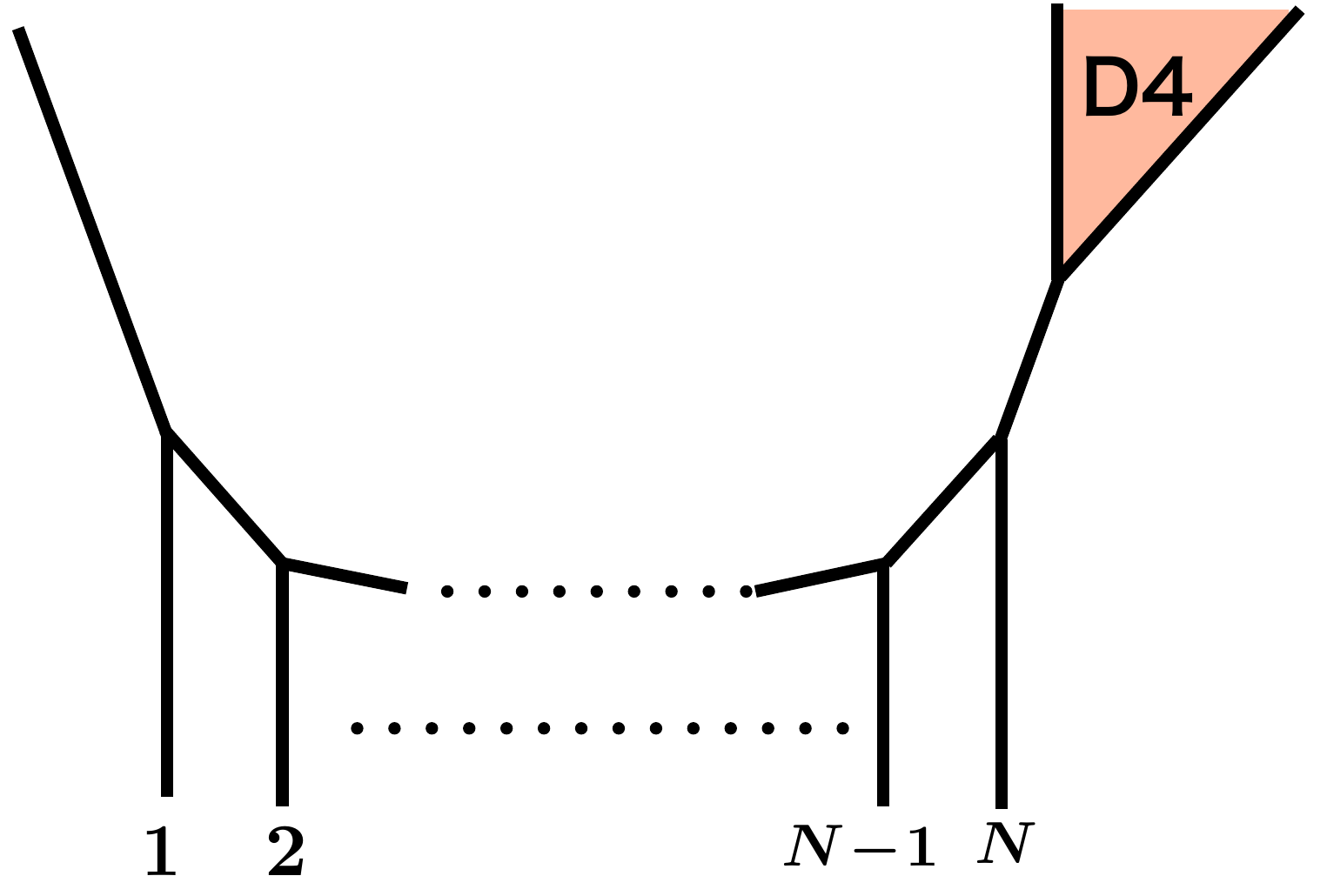}
\caption{The toric webdiagram for the moduli region ${\rm Im}\,z_0<0,\,{\rm Im}\,z_i>0,\, |{\rm Im}\,z_0| > \sum_{i=1}^{N-1} {\rm Im}\,z_i$.}
\label{fig:C2_AN}
\end{center}
\end{figure}

Among these $N+1$ regions \eqref{eq:cones1_AN}-\eqref{eq:cones3_AN} of the moduli space, there are two special regions, that is, $R_0$ and $R_N$. The former is associated with the toric diagram of figure \ref{fig:dummy_cycle}, while the latter is described by figure \ref{fig:C2_AN}.
In $R_N$, the D4-brane is wrapped on a divisor isomorphic to $\mathbb{C}^2$.
Therefore, if we take a large radii limit of ${\rm Im}\,z_0 \to -\infty,\,{\rm Im}\,z_i \to +\infty$ while keeping ${\rm Im}\,z_0\ll -\sum_{i=1}^{N-1}{\rm Im}\,z_i$, the BPS partition function of our D4-D2-D0 states should be equal to the instanton partition function on $\mathbb{C}^2$:
\begin{eqnarray}
 \mathcal{Z}_{\mathbb{C}^2}(q,Q) &=& \prod_{n=1}^\infty\frac{1}{1-q^n} = \frac{1}{\phi(q)}.
\label{eq:C2_AN}
\end{eqnarray}
On the other hand, if we take a large radii limit of ${\rm Im}\,z_I\to +\infty$ for $I=0,1,2,\cdots,N-1$, then the BPS partition function should be an instanton partition function on the divisor whose projection onto the toric base is the shaded region of figure \ref{fig:dummy_cycle}. If we neglect in this limit the effects of the dummy cycle $\beta_0$, then such an instanton partition function is expected to be given by the affine $SU(N)$ character.

\begin{figure}
\begin{center}
\includegraphics[width=13cm]{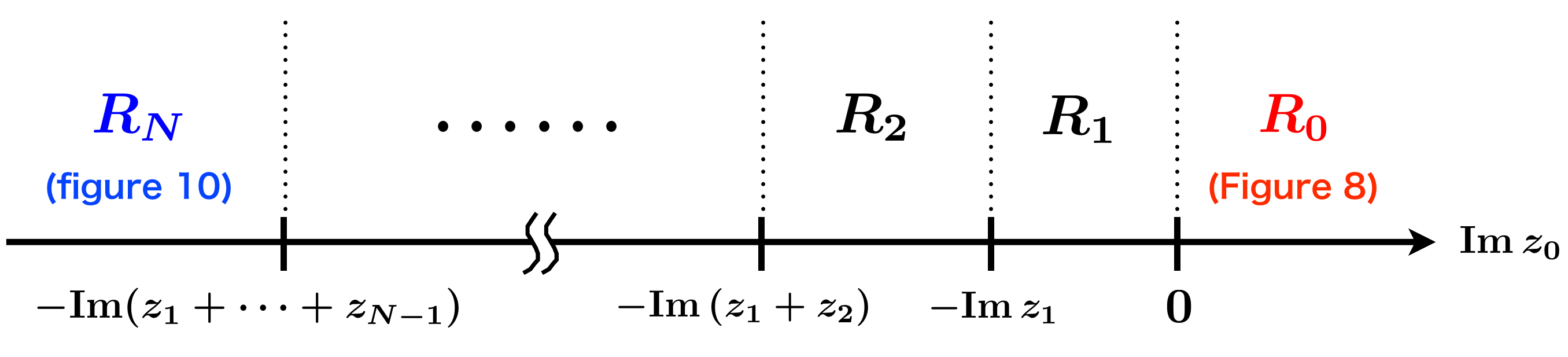}
\caption{There are $N+1$ topologically different resolutions of the Calabi-Yau three-fold, which are connected by flop transitions. In particular, two special regions $R_0$ and $R_N$ are associated with the toric diagrams of figure 8 and figure 10, respectively.}
\label{fig:multiple_topology-changes}
\end{center}
\end{figure}

These two special large radii limits can be connected by moving ${\rm Im}\,z_0$ from (i) ${\rm Im}\,z_0 = -\infty \ll -\sum_{i=1}^{N-1}{\rm Im}\,z_i$ to (ii) ${\rm Im}\,z_0 = +\infty$, while keeping ${\rm Im}\,z_1,\cdots, {\rm Im}\,z_{N-1}$ very large (See figure \ref{fig:multiple_topology-changes}). The former leads to \eqref{eq:C2_AN}, while the latter should give us the affine $SU(N)$ character \eqref{eq:instanton-A_N} after neglecting the effects of the dummy cycle $\beta_0$. Below, we will show that the wall-crossings of D4-D2-D0 states correctly interpolate these two large radii limits, and the affine $SU(N)$ character can be obtained from \eqref{eq:C2_AN} by using the wall-crossing formula \eqref{eq:wall-crossing_formula}.

\subsection{Walls of marginal stability}

As shown in subsection \ref{subsec:wall-crossing_formula}, the relevant wall-crossings are associated with separations of D2-D0 fragments from our D4-D2-D0 states:
\begin{eqnarray}
\Gamma = \mathcal{D}+k^I\beta_I - ldV \quad \to\quad (\Gamma_1 = \Gamma - \Gamma_2) \quad +\quad (\Gamma_2 = m^I\beta_I - ndV),\label{eq:D2-D0_separation_AN}
\end{eqnarray}
where the sum $I=0,\cdots,N-1$ is implicit.
What we need to know in order to use the wall-crossing formula \eqref{eq:wall-crossing_formula} are the BPS indices of D2-D0 fragments and the locations of the walls of marginal stability.

The D2-D0 indices can be read off from the Gopakumar-Vafa invariants. The non-vanishing Gopakumar-Vafa invariants for the Calabi-Yau of figure \ref{fig:dummy_cycle} are
\begin{eqnarray}
 N^\beta_0 &=& 1\qquad {\rm for}\qquad \beta= \beta_0 + \beta_1+\cdots + \beta_j, \quad 0\leq j\leq N-1,\label{eq:GV_AN1}
\\[2mm]
N^\beta_0 &=& -1\qquad {\rm for}\qquad \beta= \beta_i+ \beta_{i+1}+\cdots+\beta_{j},\quad 1\leq i\leq j\leq N-1.
\nonumber \\
\label{eq:GV_AN2}
\end{eqnarray}
The higher genus contributions again vanish.
For the derivation of these results, see appendix \ref{app:Gopakumar-Vafa}.
From the Gopakumar-Vafa invariants, we find that the D2-D0 indices on our Calabi-Yau three-fold are given by\footnote{We again omit the pure D0 index, because the separation of pure D0-branes does not give rise to any wall-crossings.}
\begin{eqnarray}
\Omega(\Gamma_2) &=& 1\qquad {\rm for}\qquad \Gamma_2 = \pm\left(\beta_0 + \beta_1+\cdots+\beta_j\right) - ndV,\quad 0\leq j\leq N-1,\label{eq:D2-D0_AN1}
\\[2mm]
\Omega(\Gamma_2) &=& -1\qquad {\rm for}\qquad \Gamma_2 = \pm\left(\beta_i+\beta_{i+1}+\cdots+\beta_j\right) - ndV,\quad 1\leq i\leq j\leq N-1.
\nonumber \\
\label{eq:D2-D0_AN2}
\end{eqnarray}

For each of the D2-D0 indices, there is a wall of marginal stability of the type of \eqref{eq:D2-D0_separation_AN}.
However, as in the analysis of section \ref{sec:affine_SU(2)}, walls for \eqref{eq:D2-D0_AN2} give rise to no change in the D4-D2-D0 indices. The reason for this is that they have the vanishing charge intersection product with our D4-D2-D0 states:
\begin{eqnarray}
 \langle \Gamma_2, \Gamma\rangle = 0.
\end{eqnarray}
Therefore, it is sufficient to consider the walls for \eqref{eq:D2-D0_AN1}.

\begin{figure}
\begin{center}
\includegraphics[width=15cm]{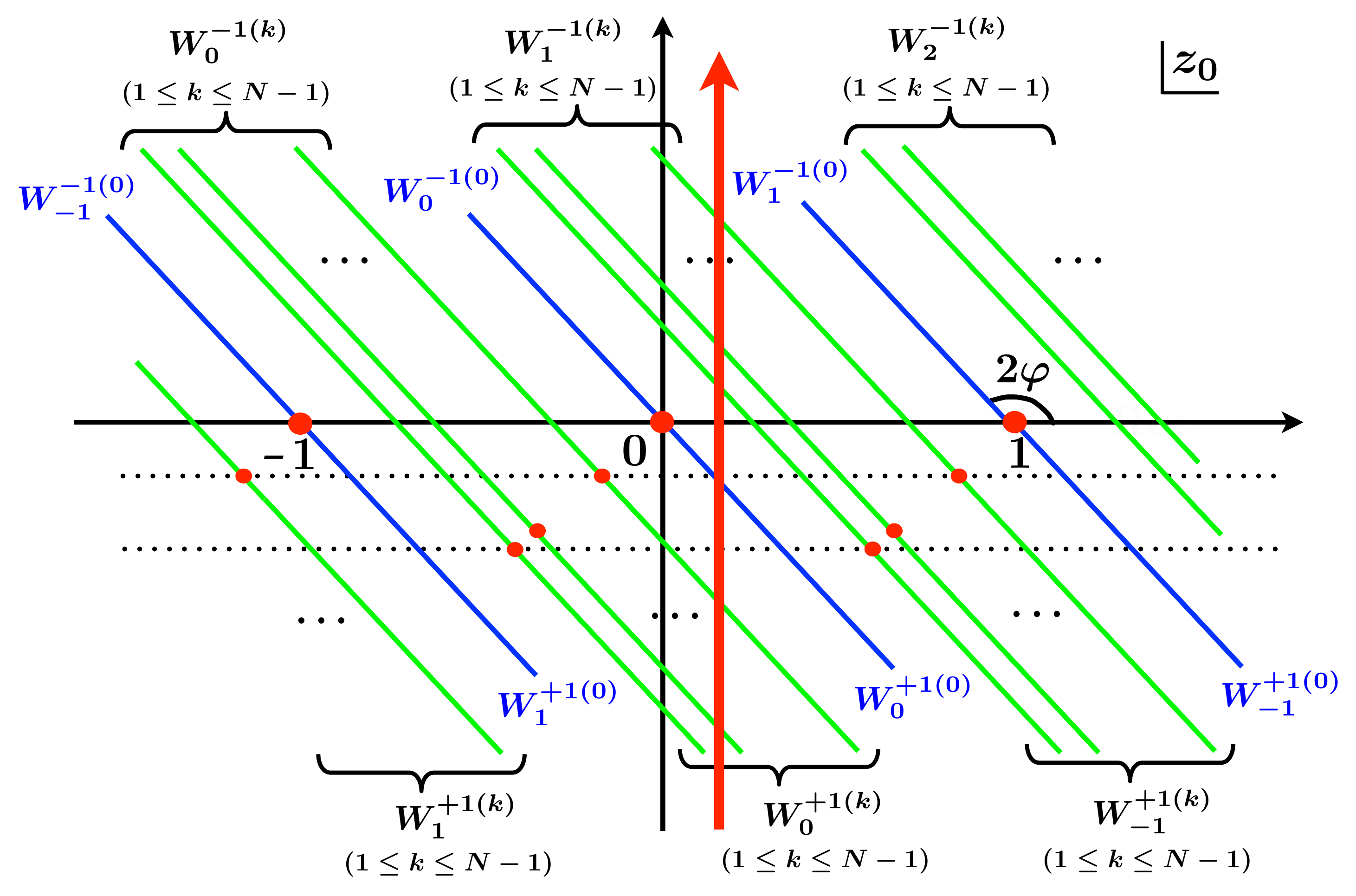}
\caption{The walls of marginal stability in $z_0$-plane with $z_i$ fixed for $i=1,\cdots,N-1$. The red dots denote singularities where some D2-D0 state becomes massless. The upper dotted line expresses ${\rm Im}\,z_0 = -{\rm Im}\,z_1$ while the lower one is ${\rm Im}\,z_0=-\sum_{i=1}^{N-1}{\rm Im}\,z_i$.}
\label{fig:walls_AN}
\end{center}
\end{figure}

We now identify the corresponding walls of marginal stability by solving $\arg Z(\Gamma_1) = \arg Z(\Gamma_2)$. Let us denote by $W^{\pm1 (k)}_n$ the wall associated with the separation of D2-D0 fragments of charge $\Gamma_2 = \pm (\beta_0+\beta_1+ \cdots + \beta_k) - ndV$. The relevant central charges for the wall $W^{\pm1(k)}_n$ are evaluated as
\begin{eqnarray}
 Z(\Gamma_1) \sim -\frac{c_4}{2}\Lambda^2e^{2i\varphi},\qquad Z(\Gamma_2) = \pm \left(z_0+z_1+\cdots+z_k\right)+ n,
\end{eqnarray}
and therefore the location of the wall $W^{\pm1(k)}_n$ is specified by
\begin{eqnarray}
 2\varphi &=& \arg[\mp\left(z_0+z_1+\cdots+z_k\right) - n].
\end{eqnarray}
We can draw the walls $W^{\pm1(k)}_n$ in $z_0$-plane with $z^i$ fixed for $i=1,\cdots,N-1$, as in figure \ref{fig:walls_AN}.
All the walls are straight lines in $z_0$-plane, with the slope of $2\varphi$. The blue and green walls express $W^{\pm1 (0)}_{N}$ and $W^{\pm1 (k)}_{N}$ for $1\leq k\leq N-1$, respectively. The red dots denote the singularities where some D2-D0 states becomes massless. The locations of the green walls depend on $z_1,\cdots,z_{N-1}$, while the blue walls are invariant under changing $z_1,\cdots,z_{N-1}$. In particular, if we keep ${\rm Im}\,z_i$ very large for $i=1,2,\cdots,N-1$, then the green walls move down in figure \ref{fig:walls_AN}.

In the next subsection, we will move the moduli parameter $z_0$ from ${\rm Im}\,z_0 = -\infty$ to ${\rm Im}\,z_0 = +\infty$ along the red arrow in figure \ref{fig:walls_AN}, and evaluate the discrete changes of the BPS partition function of our D4-D2-D0 states.

\subsection{Affine $SU(N)$ character}

Now, we come to the main result of this paper. We will here show that the affine $SU(N)$ character can be obtained from the instanton partition function on $\mathbb{C}^2$, by considering the wall-crossing of D4-D2-D0 states.

We first fix $z_1\sim z_{N-1}$ so that $0<\sum_{i=1}^{N-1}{\rm Re}\,z_i <1$ and ${\rm Im}\,z_i$ is very large for $i=1,2,\cdots,N-1$. Then, we move $z_0$ from ${\rm Im}\,z_0 = -\infty$ to ${\rm Im}\,z_0 = +\infty$ along the red arrow in figure \ref{fig:walls_AN}. Although we set ${\rm Im}\,z_1,\cdots,{\rm Im}\,z_{N-1}$ to be very large, ${\rm Im}\,z_0$ is much smaller than $-\sum_{i=1}^{N-1}{\rm Im}\,z_i$ at the bottom of the red arrow. This implies that the topology of our Calabi-Yau is described by figure \ref{fig:C2_AN} at the bottom of the arrow. On the other hand, at the top of the red arrow, our Calabi-Yau three-fold is described by figure \ref{fig:dummy_cycle}. Thus, from ${\rm Im}\,z_0=-\infty$ to ${\rm Im}\,z_0 = +\infty$ along the red arrow, we have multiple topology-changes of the Calabi-Yau (See figure \ref{fig:multiple_topology-changes}).

We start from the limit of ${\rm Im}\,z_0 = -\infty$, where the BPS partition function of the D4-D2-D0 states is given by \eqref{eq:C2_AN}. By moving ${\rm Im}\,z_0$ away from ${\rm Im}\,z_0 = -\infty$, various walls of marginal stability are crossed, and the BPS partition function changes discretely. We can evaluate the discrete changes of the BPS partition function by using the wall-crossing formula \eqref{eq:wall-crossing_formula}. Along the red arrow in figure \ref{fig:walls_AN}, the moduli cross all the walls of $W^{+1(k)}_{N}$ for $N=0,1,2,\cdots$ and $W^{-1(k)}_{N}$ for $N=1,2,3,\cdots$. Recall here that the form of \eqref{eq:wall-crossing_formula} is valid when the ordering of the central charge phases are reversed from $\arg Z(\gamma_2)<\arg Z(\Gamma_1)$ to $\arg Z(\Gamma_2)> \arg Z(\Gamma_1)$ at the wall-crossing. Therefore, if the moduli cross $W^{+1(k)}_N$ along the red arrow, we can use \eqref{eq:wall-crossing_formula} without any modification. However, if the moduli cross $W^{-1(k)}_N$ along the arrow, we have to reverse the sign of the exponent in \eqref{eq:wall-crossing_formula}. 

From the above argument, we find that the walls of $\{W^{+1(k)}_{n\geq 0}\}$ give rise to the multiplication of
\begin{eqnarray}
 \prod_{n=0}^\infty(1-q^nQ_0)(1-q^nQ_0Q_1)\cdots(1-q^nQ_0\cdots Q_{N-1})
\end{eqnarray}
to the partition function. Here, we used \eqref{eq:D2-D0_AN1} and the fact that $\langle\Gamma_2,\Gamma\rangle = +1$ for $\Gamma_2 = +(\beta_0 +\beta_1+\cdots+\beta_k)-ndV$, in the wall-crossing formula \eqref{eq:wall-crossing_formula}. On the other hand, the walls of $\{W^{-1(k)}_n\}$ give rise to the following multiplication to the partition function:
\begin{eqnarray}
 \prod_{n=1}^\infty(1-q^nQ_0^{-1})(1-q^nQ_0^{-1}Q_1^{-1})\cdots(1-q^nQ_0^{-1}\cdots Q_{N-1}^{-1}),
\end{eqnarray}
where we used \eqref{eq:D2-D0_AN1} and the fact that $\langle\Gamma_2,\Gamma\rangle = -1$ for $\Gamma_2 = -(\beta_0+\cdots+\beta_{k}) - ndV$. Note here that, the minus sign of $\langle \Gamma_2,\Gamma\rangle$ cancels the reversed sign of the exponent in \eqref{eq:wall-crossing_formula}. From these arguments, we find that the BPS partition functions $\mathcal{Z}_{\pm\infty}$ in the limit of ${\rm Im}\,z_0 = \pm\infty$ are related to each other by
\begin{eqnarray}
 \mathcal{Z}_{+\infty}(q,Q) &=& \mathcal{Z}_{-\infty}(q,Q)\prod_{k=0}^{N-1}\left[\prod_{n=0}^\infty(1-q^nQ_0\cdots Q_k)\prod_{m=1}^\infty(1-q^mQ_0^{-1}\cdots Q_k^{-1})\right]
\nonumber \\
\label{eq:relation_AN}
\end{eqnarray}
Recall here that the BPS partition function $\mathcal{Z}_{-\infty}$ should be equivalent to \eqref{eq:C2_AN}:
\begin{eqnarray}
 \mathcal{Z}_{-\infty}(q,Q) = \prod_{n=1}^\infty\frac{1}{1-q^n} = \frac{1}{\phi(q)}.
\end{eqnarray}
Then, we can use the relation \eqref{eq:relation_AN}, which is the result of the wall-crossings, to obtain the explicit expression of $\mathcal{Z}_{+\infty}$:
\begin{eqnarray}
 \mathcal{Z}_{+\infty}(q,Q)
&=& \frac{1}{\phi(q)^{N+1}}\prod_{k=0}^{N-1}\sum_{n\in\mathbb{Z}}q^{\frac{1}{2}n(n-1)}(-Q_0\cdots Q_k)^n
\nonumber \\
 &=& \frac{1}{\phi(q)^{N+1}}\!\!\!\!\sum_{n_0,\cdots,n_{N-1}\in\mathbb{Z}}\!\!(-1)^{\sum_{\ell=0}^{N-1}n_\ell}\,q^{\frac{1}{2}\sum_{i=0}^{N-1}n_i(n_i-1)}\prod_{j=0}^{N-1}Q_{j}^{\sum_{k=j}^{N-1}n_k},\label{eq:Z_infty_AN}
\end{eqnarray}
where we used the identity \eqref{eq:identity}.
This is the BPS partition function in the large radii limit of ${\rm Im}\,z_I\to +\infty$ for $I=0,\cdots,N-1$, and should be equivalent to the instanton partition function on a divisor whose projection onto the toric base is the shaded region in figure \ref{fig:dummy_cycle}. 

What we should do next is to extract the $Q_0$-independent terms from \eqref{eq:Z_infty_AN}, in order to neglect the contributions from D2-branes wrapped on the dummy cycle $\beta_0$. By imposing $n_0 = -(n_1+n_2+\cdots+n_{N-1})$ in the summation in \eqref{eq:Z_infty_AN}, we obtain 
\begin{eqnarray}
&&\left.\mathcal{Z}_{+\infty}(q,Q)\right|_{Q_0{\rm -independent}}
\nonumber \\[2mm]
&&\qquad\qquad =
 \frac{1}{\phi(q)^{N+1}}\!\!\!\!\sum_{n_1,\cdots,n_{N-1}\in\mathbb{Z}}q^{\frac{1}{2}(n_1+n_2+\cdots+n_{N-1})^2+\frac{1}{2}\sum_{i=1}^{N-1}n_i^2}\prod_{j=1}^{N-1}Q_{j}^{\sum_{k=j}^{N-1}n_k}.
\end{eqnarray}
By using different summation variables $\widetilde{n}_j = \sum_{k=j}^{N-1}n_k$, we can rewrite this as
\begin{eqnarray}
\left.\mathcal{Z}_{+\infty}(q,Q)\right|_{Q_0{\rm -independent}}
&=&
\frac{1}{\phi(q)^{N+1}}\!\!\!\!\sum_{\widetilde{n}_1,\cdots,\widetilde{n}_{N-1}\in\mathbb{Z}}q^{\sum_{i=1}^{N-1}\widetilde{n}_i^2 - \sum_{i=1}^{N-2}\widetilde{n}_i\widetilde{n}_{i+1}}\prod_{j=1}^{N-1}Q_{j}^{\widetilde{n}_j}
\nonumber \\[2mm]
&=&  \frac{q^{\frac{N+1}{24}}}{\eta(q)^2}\;\chi_{0}^{\widehat{su}(N)_1}(q,Q).
\label{eq:result_AN}
\end{eqnarray}
In the last equality, we recall the expression \eqref{eq:character-zero} of $\chi^{\widehat{su}(N)_1}_0$.
This is correctly proportional to the level one affine $SU(N)$ character! The prefactor is related to the D4-D0 degeneracy and independent of D2-brane chemical potentials $Q_i$. The $Q_i$-dependent terms are perfectly given by the character of $\widehat{su}(N)$. This result implies that the wall-crossing formula for d=4, $\mathcal{N}=2$ supersymmetric theories knows about the instantons on ALE spaces. The physical meaning of the parameter $r$ of $\chi_r^{\widehat{su}(N)_1}$ should again be related to the boundary condition at infinity, and left for future work.

\section{Discussions}
\label{sec:discussions}

In this paper, we have studied the relation between instanton partition functions on $A_{N-1}$-ALE spaces and D4-D2-D0 partition functions on toric Calabi-Yau three-folds, mainly concentrating on the wall-crossing phenomena. We have obtained the instanton partition function on $A_{N-1}$-ALE spaces from that on $\mathbb{C}^2$ via the wall-crossing formula. The result correctly proportional to the affine $SU(N)$ character, which is consistent with the work by Nakajima in \cite{Nakajima}. Our results imply that the wall-crossing formula of $d=4,\mathcal{N}=2$ supersymmetric theories knows about the relation between instanton partition functions on $A_{N-1}$-ALE space and $\mathbb{C}^2$.

We here briefly go into the further detail of the agreement between \eqref{eq:result_AN} and the instanton partition function on ALE space. Recall that the instanton partition function on $A_{N-1}$-ALE space \eqref{eq:result_AN} derived from the wall-crossings is proportional to the affine $SU(N)$ character, up to a prefactor which expresses the D4-D0 bound states without D2-branes. Such a prefactor is interpreted as regular instantons on the divisor wrapped by the D4-brane. In fact, \eqref{eq:result_AN} nicely agrees with the expression derived by Szabo in \cite{Szabo:2009vw}, including the prefactor.\footnote{See equation (4.9) in \cite{Szabo:2009vw}.} The expression in \cite{Szabo:2009vw} is the partition function of regular and fractional instantons on ALE space, which can be evaluated by the method of localization. The regular instantons contribute
\begin{eqnarray}
\left(\frac{1}{\phi(q)}\right)^{\chi(C_4)}
\end{eqnarray}
to the partition function, where $\chi(C_4)$ is the Euler characteristic of the divisor wrapped by the D4-brane. When $C_4$ is $A_{N-1}$-ALE space, $\chi(C_4) = N$. However, since we have added a dummy cycle to our Calabi-Yau three-fold and the Euler characteristic of the divisor increases by one, we now have $\chi(C_4) = N+1$ which is perfectly consistent with our prefactor of \eqref{eq:result_AN}.

We should stress here that our results give non-trivial relations between instanton partition functions of the Vafa-Witten theories on different four-manifolds. An interesting future problem is to investigate the similar relations between the Vafa-Witten partition functions on other four-manifolds via wall-crossing phenomena. 

Another future direction will be the generalization to the Calabi-Yau three-fold including compact four-cycles. When the Calabi-Yau has some compact four-cycles, the situation is drastically changed and we now have to take into account the pair creations of D4 and $\overline{\rm D4}$ in the analysis of the wall-crossings. Such a generalization will reveal some non-trivial relation between the Vafa-Witten theories on compact four-manifolds.

It is also worth studying how to interpret our results from the viewpoint of the statistical model description of BPS states. The instantons on $\mathbb{C}^2$ and $A_{N-1}$-ALE space are related to Young diagrams and orbifold partitions \cite{Dijkgraaf:2007fe}, respectively. Furthermore, the wall-crossings of D4-D2-D0 are partly understood by the two-dimensional statistical model called ``triangular partition model,'' at least for the D4-D2-D0 states on the resolved conifold \cite{Nishinaka:2011sv}. Moreover, the statistical model description of the BPS D-branes is known to be closely related to free fermions and matrix model, at least for D6-D2-D0 states \cite{Ooguri:2010yk}.\footnote{For a recent development of the matrix model description of the instanton partition function on ALE spaces, see \cite{Kimura:2011zf}.} In particular, our calculations of the closed D4-D2-D0 wall-crossings seem to be related to the free fermion and matrix model descriptions of open BPS wall-crossings studied in \cite{Sulkowski:2010eg}. It will be interesting to study their generalization to our D4-D2-D0 wall-crossings.

\acknowledgments

We would like to thank Hiroshi Itoyama, Takahiro Kubota, Takuya Okuda, Takeshi Oota and Yutaka Yoshida for many illuminating discussions, important comments and suggestions.
S.Y. was supported in part by KAKENHI 22740165.

\appendix

\section{Notation}
\label{app:notation}

In this paper, we mainly use the notation of \cite{Denef:2007vg}. Thanks to the Poincar\'e duality, the D-brane charges can be expressed by even-forms on the Calabi-Yau three-fold. To be more precise, D$p$-branes wrapped on a $p$-cycle of the Calabi-Yau $X$ is represented by an element of coholomology $H^{6-p}(X)$. So a general charge $\Gamma$ for D0-D2-D4-D6 branes on $X$ is written as
\begin{eqnarray}
 \Gamma = a + \mathcal{D} + \beta - ndV,
\end{eqnarray}
where $a\in H^{0}(X),\,\mathcal{D}\in H^2(X),\, \beta\in H^4(X),\, n\in \mathbb{Z}$ and $H^6(X)\ni dV$ is the normalized volume form so that $\int_X dV =1$. The integer $n$ denotes the charge for D0-branes localized on a Calabi-Yau $X$. We can take a basis $\beta_i$ of $H^2(X,\mathbb{Z})$ so that $\beta_i$ is Poincar\'e dual to the $i$-th two-cycle $B^i\in H_2(X)$ of the Calabi-Yau $X$. By using this basis, a general D2-brane charge $\beta\in H^2(X)$ can be expanded as $\beta = m^i\beta_i$, where $m^i$ is the charge for D2-branes wrapped on the $i$-th two-cycle. The two-form $\mathcal{D}$ represents the D4-brane charge. Although we can generally consider multiple D4-branes, in this paper we concentrate on the case of a single D4-brane. So $\mathcal{D}$ is Poincar\'e dual to the divisor wrapped by the D4-brane. If we also consider D6-branes wrapped on the whole Calabi-Yau, we should introduce $a\neq 0$.

The Dirac-Schwinger-Zwanziger intersection product of the charges are defined by
\begin{eqnarray}
\langle\Gamma,\Gamma'\rangle &=& \int_X\Gamma\wedge \check{\Gamma}',
\end{eqnarray}
where $\check{\Gamma}$ is an even-form obtained from $\Gamma$ by inverting the sign of the 2-form and 6-form. In terms of the components, this can be written as
\begin{eqnarray}
 \langle\Gamma,\Gamma'\rangle &=& (an' - a'n) + \sum_{i}(m'^i - m^i)\#(D\cap B^i)
\end{eqnarray}
where $D$ and $B^i$ are four and two-cycles dual to $\mathcal{D}$ and $\beta_i$, respectively. The quantity
\begin{eqnarray}
\langle \mathcal{D},\beta_i\rangle\; =\; \int_X\mathcal{D}\wedge \beta_i\;=\; \#(D\cap B^i)
\end{eqnarray}
 represents the intersection number of $D$ and $B^i$ in the Calabi-Yau $X$. If we consider a toric Calabi-Yau three-fold, the intersection number can be read off from the toric diagram.

The central charge of our D-brane bound states is approximately evaluated as
\begin{eqnarray}
 Z(\Gamma) &=& -\int_X\Gamma\wedge e^{-t}
\end{eqnarray}
in the large radii limit. Here $t$ is the K\"ahler two-form of the Calabi-Yau, and this quantity roughly expresses the sum of the complexified volume wrapped by the D-branes. In fact, by expanding the exponential we find
\begin{eqnarray}
 Z(\Gamma) &=& \frac{a}{6}\int_X t\wedge t\wedge t - \frac{1}{2}\int_X\mathcal{D}\wedge t\wedge t + m^i\int_X\beta_i \wedge t + n\int_X dV
\nonumber \\
&=& \frac{a}{6}\int_Xt\wedge t\wedge t - \frac{1}{2}\int_{D}t\wedge t + m^i\int_{B^i}t + n.
\end{eqnarray}
If the moduli move away from the large radii limit, we can no longer trust this expression of the central charge, due to the non-trivial $\alpha'$-corrections.

\section{BPS index}
\label{app:BPS index}

The index for BPS states in $d=4,\mathcal{N}=2$ supersymmetric theories are defined as
\begin{eqnarray}
 \Omega(\Gamma) = -\frac{1}{2}{\rm Tr}_{\mathcal{H}_\Gamma}\left[(-1)^{2J}(2J)^2\right],
\label{eq:BPS_index}
\end{eqnarray}
where $\Gamma$ denotes the electric and magnetic charges and $J$ stands for the third component of the angular momentum operator. The trace is taken over the Hilbert space of {\em one-particle states} carrying charge $\Gamma$. In $\mathcal{N}=2$ supersymmetric theories, a massive state belongs to a short or long supersymmetric multiplet. States that belong to the short multiplets are called BPS states, while the other non-BPS states form long super multiplets. A short multiplet is composed of two spin $j$ states and one state of spin $j \pm 1/2$. They can be expressed as $[j]\otimes (2[0] \oplus [\frac{1}{2}])$, where $2[0] \oplus [\frac{1}{2}]$ is called the half hyper multiplet. The long super multiplet, in turn, can be written as $[j]\otimes (2[0] \oplus [\frac{1}{2}])^2$. Since both the short and long multiplets include at least one half hyper multiplet, we can trace out it in \eqref{eq:BPS_index} to obtain
\begin{eqnarray}
 \Omega(\Gamma) = {\rm Tr}_{\mathcal{H}_\Gamma/(2[0]\oplus[\frac{1}{2}])}(-1)^{2j}.
\end{eqnarray}
Thus, the BPS index $\Omega(\Gamma)$ is regarded as the ``Witten index for the short multiplets.'' Since a long super multiplet involves the equal numbers of ``bosonic'' and ``fermionic'' short multiplets, its contribution to the index is zero. Therefore, only the BPS states can contribute to the index \eqref{eq:BPS_index}.

The BPS partition function $\mathcal{Z}$ is defined as a generating function of the BPS indices:
\begin{eqnarray}
\mathcal{Z} &=& \sum_{\Gamma}\Omega(\Gamma)e^{\Gamma},\label{def-partition-function1}
\end{eqnarray}
where $e^\Gamma$ stands for abstract Boltzmann weight satisfying $e^{\Gamma_1+\Gamma_2} = e^{\Gamma_1}e^{\Gamma_2}$. For our D4-D2-D0 states, we define
\begin{eqnarray}
q = e^{-dV},\qquad Q_i = e^{\beta_i},
\end{eqnarray}
so that $q$ and $Q_i$ are the chemical potentials for D0-branes and D2-branes wrapped on $i$-th two-cycle, respectively.

What is important is that the BPS index of D-branes wrapped on a Calabi-Yau three-fold can be seen as the Witten index of the field theory on the D-branes, if we can neglect the $\alpha'$-corrections.\footnote{Since the existence of the D-branes already breaks a half of supersymmetry, the Witten index on the D-branes counts the BPS states in the target space as long as there is no $\alpha'$-correction.} It is for this reason that the BPS partition function is equivalent to the instanton partition function on the D-branes in the field theory limit.

\section{Flop transitions}
\label{app:flop}

\begin{figure}
\begin{center}
\includegraphics[width=4cm]{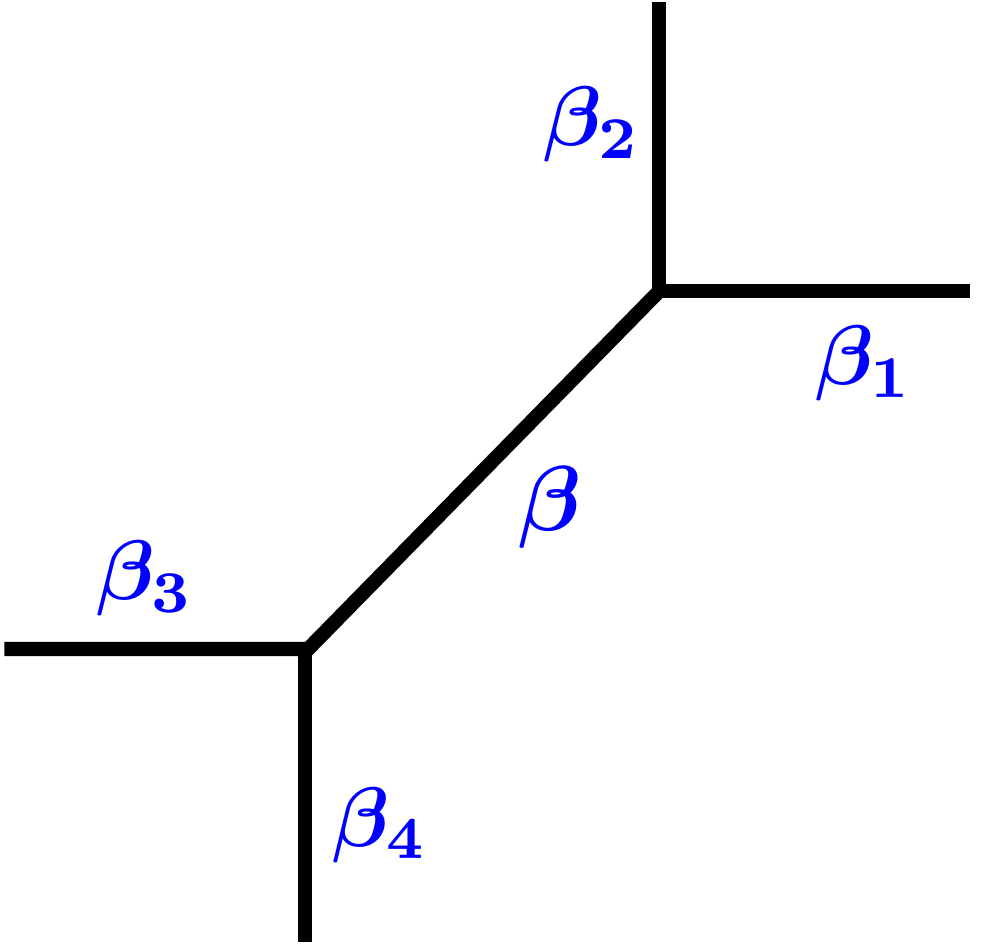}\qquad\qquad
\includegraphics[width=4cm]{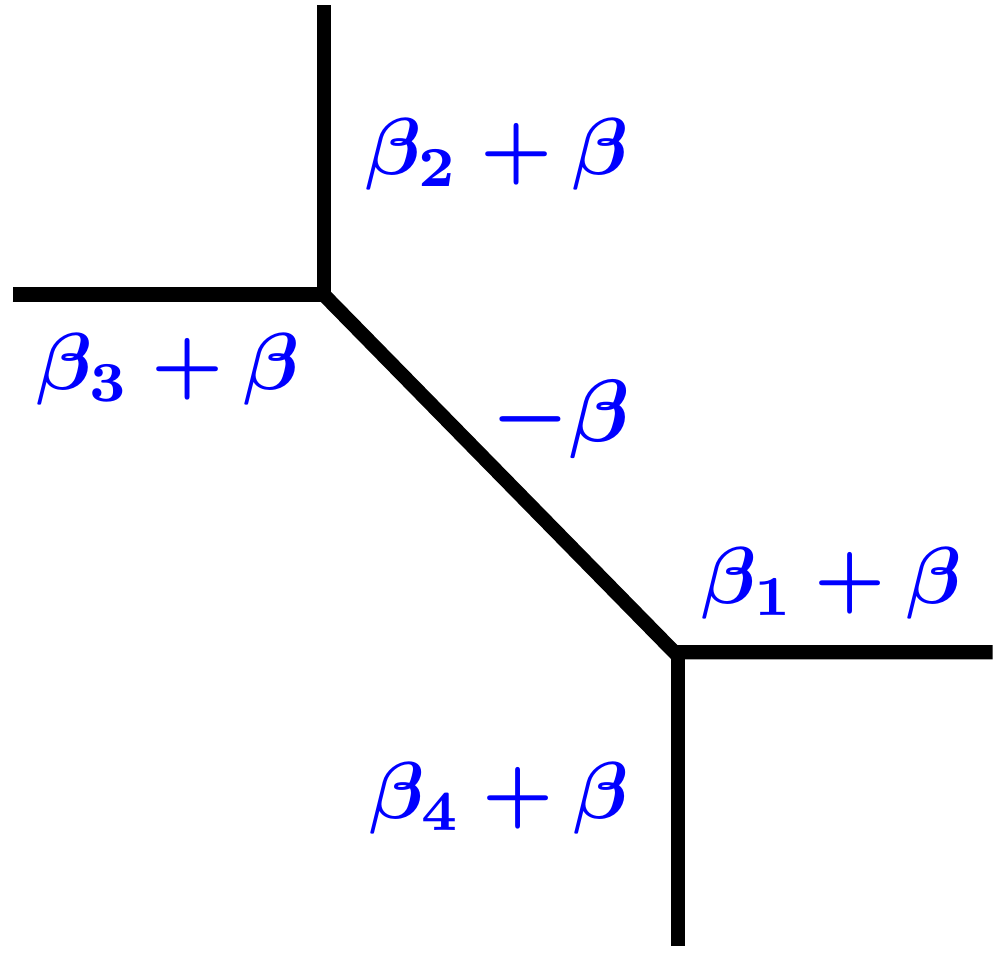}
\caption{Left: The toric webdiagram of the resolved conifold.\; Right: The toric diagram after the flop transition. The basis of the two-cycles are changed.}
\label{fig:toric_conifold}
\end{center}
\end{figure}

The flop transition is associated to a resolved conifold geometry $\mathcal{O}(-1)\oplus\mathcal{O}(-1)\to\mathbb{P}^1$. The toric webdiagram of the resolved conifold can be depicted as in the left picture of figure \ref{fig:toric_conifold}.
The internal and external lines express the compact and non-compact two-cycles in the resolved conifold. If this resolved conifold geometry is embedded into a more complicated toric Calabi-Yau three-fold, then the external edges might represent some compact two-cycles of the total Calabi-Yau three-fold.

The flop transition occurs when the two-cycle $\beta$ shrinks to the zero size. Let $z$ be the K\"ahler parameter of $\beta$ so that the K\"ahler two-form of the Calabi-Yau includes the term
\begin{eqnarray}
z\mathcal{P}
\end{eqnarray}
for a harmonic two-form $\mathcal{P}\in H^2(X)$ satisfying $\int_{X}\mathcal{P}\wedge \beta = 1,\,\int_X\mathcal{P}\wedge \beta_i = 0$. Then the size of the two-cycle $\beta$ is represented by ${\rm Im}\,z>0$. So if we set ${\rm Im}\,z$ to be zero, then we have the flop transition.\footnote{In order to keep D2-branes wrapped on $\beta$ massive, we keep ${\rm Re}\,z\not\in\mathbb{Z}$ at the flop transition. If we set ${\rm Re}\,z = {\rm Im}\,z = 0$, then we have the geometric transition rather than the flop.} If the moduli move to the region ${\rm Im}\,z<0$ after the flop, the topology of the Calabi-Yau becomes changed. In particular, the toric diagram is now depicted as in the right picture of figure \ref{fig:toric_conifold}. Now, the basis of the two-cycles are changed because the moduli move to a different K\"ahler cone of the Calabi-Yau. To be more precise, we now have a different resolution of the conifold geometry. The resolved two-cycle is now expressed by
\begin{eqnarray}
-\beta,
\end{eqnarray}
and the neighboring two-cycles are also changed as
\begin{eqnarray}
 \beta_i \to \beta_i + \beta.
\end{eqnarray}

\section{Gopakumar-Vafa invariants}
\label{app:Gopakumar-Vafa}

The Gopakumar-Vafa invariants $N^\beta_r$ physically express the BPS index of M2-branes wrapped on the two-cycle $\beta$ of a Calabi-Yau three-fold. Such an index can be seen as D2-D0 index when we compactify a spatial direction transverse to the Calabi-Yau.

On the other hand, if we compactify the temporal direction as an M-theory circle, we can relate the Gopakumar-Vafa invariants to the topological string amplitude on the Calabi-Yau three-fold, where the original wrapped M2-branes are seen as world-sheet instantons of the topological string. To be more specific, the topological string amplitude $\mathcal{F}_{\rm top}(q,Q)$ can be expanded as
\begin{eqnarray}
 \mathcal{F}_{\rm top}(q,t) &=& \sum_{\beta\in H^2(X)}\sum_{r=0}^\infty\sum_{m=1}^\infty\frac{N^\beta_r}{m}\left(q^{\frac{m}{2}} - q^{-\frac{m}{2}}\right)^{2r-2}e^{2\pi m \beta\cdot t},
\end{eqnarray} 
where $e^{2\pi t_i} = Q_i$ is the chemical potential of M2-branes wrapped on the $i$-th two-cycle of the Calabi-Yau.

We can estimate the Gopakumar-Vafa invariants by calculating the topological string partition function $\mathcal{Z}_{\rm top} = \exp \mathcal{F}_{\rm top}$. Since our Calabi-Yau is a toric Calabi-Yau three-fold expressed by figure \ref{fig:dummy_cycle}, the topological string partition function is evaluated by the topological vertex \cite{Aganagic:2003db} as
\begin{eqnarray}
\mathcal{Z}_{\rm top}(q,Q) &=& \sum_{R_0,\cdots,R_{N-1}}C_{\bullet \bullet R_0^t}\,C_{R_0\bullet R_1^t}\,C_{R_1\bullet R_2^t}\,\cdots\, C_{R_{N-1}\bullet\bullet}\,q^{\frac{1}{2}\sum_{i=1}^{N-1}\kappa(R_i)}\prod_{i=0}^{N-1}Q_i^{|R_i|},
\nonumber \\
\label{eq:topological-vertex}
\end{eqnarray}
where $|R|$ denotes the number of boxes included in the Young diagram corresponding to the representation $R$, and $q^{\frac{1}{2}\sum_{i=1}^{N-1}\kappa(R_i)}$ comes from the non-standard framings at $\beta_1,\cdots,\beta_{N-1}$.
Since \eqref{eq:topological-vertex} only includes $C_{R\bullet\bullet}$ and $C_{RR'\bullet}$, we can rewrite it in terms of the Schur functions:
\begin{eqnarray}
 \mathcal{Z}_{\rm top}(q,Q) &=& \sum_{R_0,\cdots,R_{N-1}}Q_0^{|R_0|}S_{R_0^t}(q^\rho)S_{R_{N-1}}(q^\rho) \prod_{j=1}^{N-1}Q^{|R_j|}_j \sum_{r}S_{R_{j-1}/r}(q^{\rho})S_{R_j/r}(q^{\rho})
\label{eq:iteration1}
\\
&=& \sum_{R_0,\cdots,R_{N-1}}\sum_{r_1,\cdots,r_{N-1}}Q_0^{|R_0|}S_{R_0^t}(q^\rho)S_{R_{0}/r_1}(q^\rho) \prod_{j=1}^{N-1}Q^{|R_j|}_j S_{R_{j}/r_j}(q^{\rho})S_{R_j/r_{j+1}}(q^{\rho}),
\nonumber \\
\end{eqnarray}
where $q^\rho$ is a short-hand notation of $\{x_i = q^{i-\frac{1}{2}}\}$ and $r_N = \bullet$. In fact, what we need to calculate here is similar to those in \cite{2003math.....11237Z, Eguchi:2003it}.
By using an identity $Q^{|R|}S_{R/r}(q^\rho) = Q^{|r|}S_{R/r}(Qq^\rho)$, we can rewrite this as
\begin{eqnarray}
\mathcal{Z}_{\rm top} (q,Q) &=& \sum_{R_0,\cdots,R_{N-1}}\sum_{r_1,\cdots,r_{N-1}}S_{R_0^t}(Q_0q^\rho)S_{R_{0}/r_1}(q^\rho) \prod_{j=1}^{N-1}Q^{|r_j|}_j S_{R_{j}/r_j}(Q_jq^{\rho})S_{R_j/r_{j+1}}(q^{\rho}).
\nonumber \\
\label{eq:top-schur1}
\end{eqnarray}

Now we use the identities of the skew Schur functions
\begin{eqnarray}
\sum_{R}S_{R^t/r}(x)S_{R/r'}(y) &=& \prod_{a,b=1}^\infty(1-x_ay_b)\sum_{R}S_{r'^t/R}(x)S_{r^t/R^t}(y)
\\
\sum_{R}S_{R/r}(x)S_{R/r'}(y) &=& \prod_{a,b=1}^{\infty}(1-x_ay_b)^{-1}\sum_{R}S_{r'/R}(x)S_{r/R}(y),
\end{eqnarray}
to obtain
\begin{eqnarray}
 \sum_{R_0}S_{R_0^t}(Q_0q^\rho)S_{R_0/r_1}(q^\rho) &=& Q_0^{|r_1|}S_{r_1^t}(q^\rho)\prod_{n=1}^\infty(1-Q_0q^n)^n
\end{eqnarray}
and
\begin{eqnarray}
\sum_{R_1,\cdots,R_{N-1}}\prod_{j=1}^{N-1}Q_{j}^{|r_j|}S_{R_j/r_j}(Q_jq^{\rho})S_{R_j/r_{j+1}}(q^\rho)
&=&  \prod_{j=1}^{N-1}\sum_{R}Q^{|r_j|}_j S_{R/r_j}(Q_jq^\rho)S_{R/r_{j+1}}(q^\rho)
\nonumber \\
&=&  \prod_{j=1}^{N-1}\sum_{R}Q^{|r_{j+1}|}_j S_{r_j/R}(Q_jq^\rho)S_{r_{j+1}/R}(q^\rho)
\nonumber \\
&&\qquad\qquad\times \prod_{n=1}^\infty(1-Q_jq^{n})^{-n}.\label{eq:expression1}
\end{eqnarray}
Here, it follows from $r_{N} = \bullet$ that \eqref{eq:expression1} is rewritten as
\begin{eqnarray}
 S_{r_{N-1}}(Q_{N-1}q^\rho)\prod_{j=1}^{N-2}\sum_{R}Q_j^{|r_j|}S_{r_j/R}(q^\rho)S_{r_{j+1}/R}(Q_jq^\rho)\prod_{n=1}^\infty(1-Q_jq^n)^n
\end{eqnarray}
By combining these, \eqref{eq:top-schur1} can be written as
\begin{eqnarray}
\mathcal{Z}_{\rm top}(q,Q) &=& \prod_{n=1}^{\infty}(1-Q_0q^n)^n\prod_{i=1}^{N-1}\prod_{m=1}^\infty(1-Q_iq^{m})^{-m}
\nonumber \\
&&\times\sum_{R_0,\cdots,R_{N-2}}\Biggl[Q_0^{|R_0|}S_{R_0^t}(q^\rho)S_{R_{N-2}}(Q_{N-1}q^\rho)
\nonumber \\
&& \qquad\qquad\times\prod_{j=1}^{N-2}\sum_{r}Q_j^{|R_{j}|}S_{R_{j-1}/r}(Q_jq^\rho)S_{R_{j}/r}(q^\rho)\Biggr],\label{eq:iteration2}
\nonumber \\
\end{eqnarray}
where we redefine the summation variables as $R_{i-1} := r_i$ and $r := R$.

In summary, the above calculation rewrite \eqref{eq:iteration1} as \eqref{eq:iteration2}. We can repeat this argument to obtain
\begin{eqnarray}
 \mathcal{Z}_{\rm top}(q,Q) &=& \prod_{n=1}^\infty\prod_{j=0}^{N-1}(1-Q_0\cdots Q_{j}q^{n})^n\prod_{m=1}^\infty\prod_{1\leq i\leq j\leq N-1}(1-Q_i\cdots Q_j q^n)^{-n}.
\end{eqnarray}
From this result, we find that $\mathcal{F}_{\rm top} = \log \mathcal{Z}_{\rm top}$ can be evaluated as
\begin{eqnarray}
 \mathcal{F}_{\rm top}(q,Q) &=& \sum_{j=0}^{N-1}\sum_{n=1}^\infty n\log(1-Q_0\cdots Q_jq^n)
\nonumber \\
&& \qquad - \sum_{1\leq i\leq j\leq N-1}\sum_{n=1}^\infty n\log(1-Q_i\cdots Q_jq^n).
\end{eqnarray}
By using an identity
\begin{eqnarray}
 \sum_{n=1}^\infty n\log (1-Qq^n) &=& \sum_{n=1}^\infty\sum_{m=1}^\infty \frac{Q^m}{m}n(q^{m})^n = \sum_{m=1}^\infty\frac{Q^m}{m}(q^{\frac{m}{2}}-q^{-\frac{m}{2}})^{-2},
\end{eqnarray}
we finally obtain
\begin{eqnarray}
 \mathcal{F}_{\rm top}(q,Q) &=& \sum_{j=0}^{N-1}\sum_{m=1}^\infty\frac{1}{m}(q^{\frac{m}{2}}- q^{-\frac{m}{2}})^{-2}(Q_0\cdots Q_j)^m
\nonumber \\
&& \qquad + \sum_{1\leq i\leq j\leq N-1}\sum_{m=1}^\infty \frac{-1}{m}(q^{\frac{m}{2}}- q^{-\frac{m}{2}})^{-2}(Q_i\cdots Q_j)^{m}.
\end{eqnarray}
This implies that the non-vanishing Gopakumar-Vafa invariants of the Calabi-Yau described by figure \ref{fig:dummy_cycle} are
\begin{eqnarray}
 N^\beta_0 &=& 1 \qquad {\rm for} \qquad \beta = \beta_0+\beta_1+\cdots+\beta_j,\quad 0\leq j\leq N-1, 
\\[2mm]
N^\beta_0 &=& -1\qquad {\rm for}\qquad \beta = \beta_i+\beta_{i+1}+\cdots+\beta_{j},\quad 1\leq i\leq j\leq N-1.
\end{eqnarray}

The flop invariance of the topological vertex is shown in \cite{Iqbal:2004ne, Konishi:2006ev}.\footnote{For the flop invariance of the refined topological vertex, see \cite{Taki:2008hb}}

\section{Character of affine $SU(N)$ algebra}
\label{app:character}

We here briefly summarize the basic facts of affine $SU(N)$ character. We mainly use the notation of \cite{Kac-book} and the appendix of \cite{Dijkgraaf:2007fe}. The character of $\widehat{su}(N)$ for an integrable highest-weight module $L(\Lambda)$ of the highest weight $\Lambda$ is defined by
\begin{eqnarray}
 {\rm ch}_{L(\Lambda)} := \sum_{\lambda \in P(\Lambda)}{\rm mult}_{L(\Lambda)}(\lambda)\;e^\lambda,
\end{eqnarray}
 where $P(\Lambda)$ is the set of weights of $L(\Lambda)$, and $e^\lambda$ stands for the formal exponential satisfying $e^{\lambda_1+\lambda_2} = e^{\lambda_1}e^{\lambda_2}$. The multiplicity ${\rm mult}_{V}(\lambda)$ is the dimension of $V_\lambda$ where $V = \oplus_{\lambda \in \mathfrak{h}^*}V_\lambda$ is the weight decomposition of $V$. The normalized character is defined by
\begin{eqnarray}
 \chi_\Lambda = e^{-m_{\Lambda}\delta}\,{\rm ch}_{L(\Lambda)},
\end{eqnarray}
where the modular anomaly is given by
$
 m_\Lambda = \frac{|\lambda + \rho|^2}{2(k+h^\vee)} -\frac{|\rho|^2}{2h^\vee}.
$
Let us denote by $\mathcal{Q}\hspace{-.55em}\raisebox{2.5ex}{}^{\circ}\hspace{.1em}$ the finite part of the root lattice. By using this, the classical theta function is defined as
\begin{eqnarray}
\Theta_\lambda = e^{k\Lambda_0}\sum_{\gamma\in \mathcal{Q}\hspace{-.45em}\raisebox{1.9ex}{}^{\circ}\hspace{.1em} +\overline{\lambda}/k}e^{-\frac{k}{2}|\gamma|^2\delta + k\gamma},
\end{eqnarray}
and the string function is also defined by
\begin{eqnarray}
 c^\Lambda_\lambda = e^{-(m_\Lambda-\frac{|\lambda|^2}{2k})\delta}\sum_{n=0}^\infty{\rm mult}_{L(\Lambda)}(\lambda-n\delta)e^{-n\delta}.
\end{eqnarray}
In terms of these two functions, the normalized character $\chi_\Lambda$ for an integral weight $\Lambda$ of level $k$ can be rewritten as
\begin{eqnarray}
\mathcal{\chi}_\Lambda = \sum_{\lambda \in P^{k}\;{\rm mod}\;(k \mathcal{Q}\hspace{-.45em}\raisebox{1.9ex}{}^{\circ}\hspace{.1em} + \mathbb{C}\delta)}c^\Lambda_\lambda \Theta_\lambda.
\end{eqnarray}
Here $P^k$ is the set of level $k$ integral weights.

In the case of $k=1$, it is well-known that the only non-vanishing string function is
\begin{eqnarray}
c^\Lambda_\Lambda = \eta(e^{-\delta})^{-(N-1)},
\end{eqnarray}
up to equivalence, which implies that the normalized character is simply written as
\begin{eqnarray}
 \chi_\Lambda^{\widehat{su}(N)_1} = \frac{1}{\eta(e^{-\delta})^{N-1}}\;\Theta_\Lambda^{\widehat{su}(N)_1}.
\end{eqnarray}
Here $\eta(e^{-\delta}) = e^{\delta/24}\prod_{n=1}^\infty(1-e^{-n\delta})^{-1}$. Now, let $\mathfrak{h}$ be the Cartan subalgebra whose dual space is decomposed as $\mathfrak{h}^* = \mathfrak{h}\hspace{-.4em}\raisebox{2.5ex}{}^{\circ}{}^* \oplus \mathbb{C}\Lambda_0 \oplus \mathbb{C}\delta$. By expressing $v\in\mathfrak{h}^*$ as $v = z + 2\pi i(-\tau \Lambda_0 + u\delta)$ with $z\in \mathfrak{h}\hspace{-.4em}\raisebox{2.5ex}{}^{\circ}{}^*$ and $\tau, u \in \mathbb{C}$, we can write the theta function as
\begin{eqnarray}
 \Theta^{\widehat{su}(N)_1}_\Lambda (\tau, z, u) &=& e^{2\pi i k u}\sum_{\gamma\in  \mathcal{Q}\hspace{-.45em}\raisebox{1.9ex}{}^{\circ}\hspace{.1em} + \overline{\Lambda}}q^{\frac{1}{2}|\gamma|^2}e^{(\gamma|z)} = e^{2\pi i k u}\sum_{\alpha \in  \mathcal{Q}\hspace{-.45em}\raisebox{1.9ex}{}^{\circ}\hspace{.1em}}q^{\frac{1}{2}|\alpha + \overline{\Lambda}|^2}e^{(\alpha + \overline{\Lambda}|z)},
\end{eqnarray}
where $q = e^{2\pi i \tau}$. In this paper, we only consider the value at the special point $u=0$.

There are $N$ level one weights $\Lambda_r$ for $r=0,1,2,\cdots,N-1$ which satisfy $(\Lambda_r|\alpha_j) = \delta_{rj}$ for $j=1,2,\cdots,N-1$ and $(\Lambda_r|\delta)=1$, and therefore we have $N$ level one characters labeled by $\{\Lambda_r\}$. However, since the classical theta function satisfies $\Theta_\Lambda = \Theta_{\Lambda +z +a\delta}$ for $z\in \mathcal{Q}\hspace{-.55em}\raisebox{2.5ex}{}^{\circ}\hspace{.2em}$ and $a\in \mathbb{C}$, we find that $\chi^{\widehat{su}(N)_1}_{\Lambda_r} = \chi^{\widehat{su}(N)_1}_{r\Lambda_1}$. In fact, we can show that $\Lambda_ r - r\Lambda_1 \in \mathcal{Q}\hspace{-.55em}\raisebox{2.5ex}{}^{\circ}\hspace{.2em}$ for affine $SU(N)$ algebra. Hereafter, we write $\chi_{\Lambda_r}^{\widehat{su}(N)_1} = \chi^{\widehat{su}(N)_1}_{r\Lambda_1}$ simply as $\chi_{r}^{\widehat{su}(N)_1}$. From the concrete expression of $\Lambda_1 = \sum_{j=1}^{N-1}\frac{N-j}{N}\alpha_j$ and $\alpha = \sum_{i} n_i\alpha_i$ for $n_i\in\mathbb{Z}$, it follows that
\begin{eqnarray}
\frac{1}{2}|\alpha + r\Lambda_1|^2 &=& \sum_{i=1}^{N-1} n_i^2 - \sum_{i=1}^{N-2}n_in_{i+1} + rn_1 + \frac{r^2}{2}\frac{N-1}{N},
\\
(\alpha + r\Lambda_1 | z) &=& \sum_{i=1}^{N-1}2n_iz_i - \sum_{i=1}^{N-2}(n_iz_{i+1} + z_i n_{i+1}) + r\frac{N-j}{N}z_j,
\end{eqnarray}
where we expand $z\in \mathfrak{h}\hspace{-.4em}\raisebox{2.5ex}{}^{\circ}{}^*$  as $z = \sum z_i \alpha_i$ for $z_i\in\mathbb{C}$.
Therefore, by defining the following variables
\begin{eqnarray}
&& Q_1 = e^{2z_1 - z_2},\quad Q_{N-1}=e^{2z_{N-1}-z_{N-2}},
\nonumber \\[1mm]
&& Q_i = e^{2z_{i}-z_{i-1} - z_{i+1}} \quad {\rm for}\quad i=2,3,\cdots,N-2,
\end{eqnarray}
the normalized character can be rewritten as
\begin{eqnarray}
 \chi_{r}^{\widehat{su}(N)_1}(q,Q) &=& \frac{1}{\eta(q)^{N-1}}\Theta_r^{\widehat{su}(N)_1}(q,Q),
\label{eq:character}
\end{eqnarray}
where
\begin{eqnarray}
\Theta_r^{\widehat{su}(N)_1}(q,Q) &\equiv& \Theta_{r\Lambda_1}^{\widehat{su}(N)_1}(\tau,z,0)
\nonumber \\
 &=&  \sum_{n_1,\cdots,n_{N-1}\in\mathbb{Z}}q^{\sum_{i=1}^{N-1}n_i^2 - \sum_{i=1}^{N-2}n_in_{i+1} + rn_1 + \frac{r^2}{2}\frac{N-1}{N}}\prod_{j=1}^{N-1}Q_j^{n_j + r\frac{N-j}{N}}.
\label{eq:Theta_function}
\end{eqnarray}
In particular, for $r=0$, we obtain
\begin{eqnarray}
 \chi_0^{\widehat{su}(N)_1}(q,Q) &=& \frac{1}{\eta(q)^{N-1}}\sum_{n_1,\cdots,n_{N-1}\in\mathbb{Z}}q^{\sum_{i=1}^{N-1}n_i^2 - \sum_{i=1}^{N-2}n_in_{i+1}}\prod_{j=1}^{N-1}Q_j^{n_j}.
\label{eq:character-zero}
\end{eqnarray}

\bibliography{ref}

\end{document}